\documentclass[10pt]{article}
\usepackage{amsmath}
\usepackage{graphicx}
\usepackage{amsfonts}
\usepackage{amssymb}
\usepackage{epsf}
\usepackage{latexsym}

\def\80{\hspace{0.8in}}

\def\nm{\mbox{m}}

\newcommand{\be}{\begin{enumerate}}
\newcommand{\ee}{\end{enumerate}}
\newcommand{\bi}{\begin{itemize}}
\newcommand{\ei}{\end{itemize}}
\newcommand{\bd}{\begin{description}}
\newcommand{\ed}{\end{description}}

\def\beq{\begin{equation}}
\def\eeq{\end{equation}}
\def\bea{\begin{eqnarray}}
\def\eea{\end{eqnarray}}
\def\foo{\footnote}
\def\hat{\widehat}
\def\tilde{\widetilde}

\def\pa{\partial}
\def\d{\textrm{d}}

\def\ttB{\mbox{\tt B}}
\def\ttC{\mbox{\tt C}}

\def\mj{\mbox{j}}

\def\mn{\mbox{n}}

\def\mp{\mbox{p}} 

\def\nm{\mbox{m}}
\def\nj{\mbox{j}}

\def\nr{\mbox{r}}
\def\ms{\mbox{s}}

\def\mH{\mbox{H}} 
\def\mI{\mbox{I}}
\def\mJ{\mbox{J}}
\def\nJ{\mbox{J}}

\def\mL{\mbox{L}}

\def\mN{\mbox{N}}

\def\mR{\mbox{R}}

\def\sd{\mbox{\scriptsize d}}

\def\si{\mbox{\scriptsize i}}
\def\sj{\mbox{\scriptsize j}} 
\def\sk{\mbox{\scriptsize k}}
\def\sll{\mbox{\scriptsize l}}  
\def\sm{\mbox{\scriptsize m}}
\def\sn{\mbox{\scriptsize n}} 
 
\def\sp{\mbox{\scriptsize p}}

\def\sr{\mbox{\scriptsize r}}
\def\sss{\mbox{\scriptsize s}}
\def\su{\mbox{\scriptsize u}}
\def\sv{\mbox{\scriptsize v}}

\def\sJ{\mbox{\scriptsize J}}
\def\sK{\mbox{\scriptsize K}}

\def\sN{\mbox{\scriptsize N}}

\def\sR{\mbox{\scriptsize R}}

\def\barp{\bar{p}}
\def\barq{\bar{q}}
\def\barr{\bar{r}}
\def\eph(B){\mbox{\scriptsize emergent(LMB)}}

\def\tj{\mbox{\tiny j}}
\def\tk{\mbox{\tiny k}}

\def\tm{\mbox{\tiny m}}

\def\tp{\mbox{\tiny p}}

\def\tJ{\mbox{\tiny J}}
\def\tK{\mbox{\tiny K}}

\def\fA{\mbox{\sffamily A}}

\def\fE{\mbox{\sffamily E}}
\def\fF{\mbox{\sffamily F}}

\def\fH{\mbox{\sffamily H}}

\def\fQ{\mbox{\sffamily Q}}

\def\fS{\mbox{\sffamily S}}
\def\fT{\mbox{\sffamily T}}
\def\fU{\mbox{\sffamily U}}
\def\fV{\mbox{\sffamily V}}
\def\fW{\mbox{\sffamily W}}

\def\sfA{\mbox{\sffamily{\scriptsize A}}}
\def\sfB{\mbox{\sffamily{\scriptsize B}}}
\def\sfC{\mbox{\sffamily{\scriptsize C}}}

\def\sfF{\mbox{\sffamily{\scriptsize F}}}

\def\tfA{\mbox{\sffamily{\tiny A}}}

\def\tip{\tilde{p}}
\def\tiq{\tilde{q}}

\begin{document}
\begin{titlepage}
\vspace{.7in}
\begin{center}
 
\vspace{2in} 

\LARGE{\bf TRIANGLELAND. II. QUANTUM MECHANICS}

\vspace{0.1in}

{\bf OF PURE SHAPE}\normalsize

\vspace{.4in}

\large{\bf Edward Anderson}$^{1}$

\vspace{.2in}

\large{\em Peterhouse, Cambridge CB2 1RD } and \normalsize 

\vspace{.2in}

\large{\em DAMTP, Centre for Mathematical Sciences, Wilberforce Road, Cambridge CB3 OWA.}

\end{center}

\begin{abstract}

Relational particle models are of value in the absolute versus relative motion debate. 
They are also analogous to the dynamical formulation of general relativity, and as such are useful for investigating 
conceptual strategies proposed for resolving the problem of time in quantum general relativity.  
Moreover, to date there are few explicit examples of these at the quantum level.  
In this paper I exploit recent geometrical and classical dynamics work to provide such a study based on  
reduced quantization in the case of pure shape (no scale) in 2-d for 3 particles (triangleland) with 
multiple harmonic oscillator type potentials.
I explore solutions for these making use of exact, asymptotic, perturbative and numerical methods.   
An analogy to the mathematics of the linear rigid rotor in a background electric field is useful 
throughout.  
I argue that further relational models are accessible by the methods used in this paper, and for 
specific uses of the models covered by this paper in the investigation of the problem of time (and 
other conceptual and technical issues) in quantum general relativity. 

\end{abstract}

\vspace{1in}


\mbox{ }

\vspace{3in}

\noindent$^1$ ea212@cam.ac.uk

\end{titlepage}

\section{Introduction}

In Euclidean relational particle mechanics \cite{ERPM, B94I, EOT, Paris, 06I, 06II, TriCl, 
08I, Cones}, only relative times, relative angles and relative separations are meaningful.  
While, in similarity relational particle mechanics \cite{SRPM, Paris, 06II, TriCl, FORD, AF09}, only 
relative times, relative angles and ratios of relative separations are meaningful.  
These mechanics are valuable models as regards the absolute versus relative motion debate \cite{AORM}.  
It is then of interest what structure one gets when one quantizes such theories.   
There has not been much work on this to date as regards nontrivial explicit examples [\cite{Smolin} 
concerns formal quantum constraints, \cite{Rovelli} is a toy of geometrical quantization, \cite{BS89, 
06I} is a semiclassical toy (but with only explicit examples for Euclidean relational particle 
mechanics in 1-d) and some simple solutions of the Dirac quantization scheme \cite{06I}].    
This paper goes further than these works in being the first quantum treatment of explicit nontrivial 
examples of similarity relational particle mechanics, while \cite{08III} goes further, rather, by 
considering more complicated explicit examples of Euclidean relational particle mechanics than 
those considered before.  
We get there using our recent understanding of reduced configuration spaces \cite{FORD} as a path to 
quantization.

In investigating conceptual strategies for the Problem of Time \cite{Battelle, DeWitt, K81, 
PW83, K91, K92, K99, EOT, Kieferbook, Smolin08} in Quantum Gravity, relational particle 
mechanics are useful analogues of GR \cite{K92, B94I, B94II, EOT, Kieferbook, 06II}; 
I view this as major long-term motivation for the current series of papers.     
This notorious problem occurs because `time' takes a different meaning in each of GR and 
ordinary Quantum Theory.  
This incompatibility underscores a number of problems with trying to replace these two branches with a 
single framework in situations in which the premises of both apply, namely in black holes and in the 
very early universe.  
One facet of the Problem of Time that shows up in attempting canonical quantization is that the lack of 
linear momentum dependence of the GR Hamiltonian constraint leads to a frozen (i.e. timeless, or 
stationary) quantum equation for the universe arises therein: the quantum counterpart of (I.11) is 
the Wheeler--DeWitt equation
\beq
\hat{\cal H}\Psi \equiv 
- {\hbar^2}
`
\left\{
\frac{1}{\sqrt{{\cal M}}}\frac{\delta}{\delta h^{{\mu\nu}}}
\left\{
\sqrt{{\cal M}}{\cal N}^{\mu\nu\rho\sigma}\frac{\delta}{\delta h^{{\rho\sigma}}}
\right\} - \xi \,\mbox{Ric}({\cal M})\right\}\mbox{'}\Psi 
-  \sqrt{h}\{\mbox{Ric}(h) - 2\Lambda\}\Psi  + \hat{\cal H}_{\mbox{\scriptsize matter}}\Psi = 0
\label{WDE} \mbox{ } ,   
\eeq
where $\Psi$ is the wavefunction of the universe.  


Before further consideration of the Problem of Time, a detour is first required about other features 
of this equation and of the toy models of it that are the subject of this paper.  
The inverted commas indicate that the Wheeler-DeWitt equation has, in addition to the Problem of Time, 
various technical problems, including 

\noindent 
A)  regularization problems -- not at all straightforward for an 
equation for a theory of an infinite number of degrees of freedom in the absense of background structure, 
while the mathematical meaningfulness of functional differential equations is open to question.
I emphasize that this is not an issue in this paper's toy models as these are for a finite number 
of degrees of freedom, similarly to the situation in minisuperspace quantum cosmology.  

\noindent 
B) There are operator-ordering issues, which this paper's toy models do exhibit an analogue of.  
To view the analogy, let ${\cal Q}_{\sfA}$ be general coordinates (spanning spatial indices and 
either particle labels or field species along with spatial dependence) with a corresponding 
configuration space metric ${\cal M}_{\sfA\sfB}$ with inverse ${\cal N}^{\sfA\sfB}$ and determinant 
${\cal M}$ (perhaps merely at the formal level). 
Then let $\frac{\nabla}{\nabla{\cal Q}_{\tfA}}$ be a partial derivative for finite systems or (perhaps 
merely formally) a functional derivative for infinite systems.  
Then the {\it Laplacian ordering} for the classical combination of configurations and their momenta 
${\cal N}^{\sfA\sfB}({\cal Q}^{\sfC}){\cal P}_{\sfA}{\cal P}_{\sfB}$ is  
\beq
D^2 = \frac{1}{\sqrt{{\cal M}}} \frac{\nabla}{\nabla {\cal Q}_{\sfA}}
\left\{
\sqrt{{\cal M}}{\cal N}^{\sfA\sfB}\frac{\nabla}{\nabla{\cal Q}_{\sfB}}
\right\} \mbox{ } , 
\eeq  
which has the desirable property of being independent of coordinate choice on the configuration space 
\cite{DeWitt57}. 
This property is not, however, unique to this ordering: one can include a Ricci scalar 
curvature term so as to have $D^2 - \xi\,\mbox{Ric}({\cal M})$ \cite{DeWitt57, HP86CZ, Oporder}. 
There is then a unique conformally-invariant choice \cite{Oporder, Banal} among these orderings.   
The conformal invariance here corresponds to retaining at the quantum level the banal conformal 
invariance that is obvious and natural in the relational action at the classical level \cite{08I, Banal}.
This ordering is 
\beq
{\cal D}^2 = \frac{1}{\sqrt{{\cal M}}} \frac{\nabla}{\nabla {\cal Q}_{\sfA}}
\left\{
\sqrt{{\cal M}}{\cal N}^{\sfA\sfB}\frac{\nabla}{\nabla{\cal Q}_{\sfB}}
\right\} 
- \frac{k - 2}{4\{k - 1\}}\mbox{Ric}({\cal M}) \mbox{ } , 
\eeq 
where $k$ is the configuration space dimension.
Furthermore, for this to be conformal, it is required that $\Psi$ itself transforms in general tensorially  
under the conformal transformation \cite{Wald}:\footnote{This paper's models, like minisuperspace, only   
require the finite configuration space dimension $k$ case of these equations.}
\beq
\Psi \longrightarrow \Psi_{\Omega} = \Omega^{\frac{2 - k}{2}}\Psi \mbox{ } .
\eeq
Moreover, this paper focusses on a model with a 2-d configuration space, for which the conformal 
$\xi = \{k - 2\}/4\{k - 1\}$ collapses to 
zero, so that  Laplacian ordering and conformally invariant wavefunctions suffice (but almost all other relational particle 
mechanics models, 
such as \cite{08III}, have configuration space dimension $\geq 3$ for which this subtlety is required).  
Finally, if one sends $\fH\Psi = \fE\Psi$ to ${\fH}_{\Omega}{\Psi}_{\Omega} = 
{\fE}_{\Omega}{\Psi}_{\Omega} = \{\fE/\Omega^2\}\Psi_{\Omega}$, one has now an eigenvalue problem with 
a weight function $\Omega^{-2}$

\noindent which then appears in the inner product: 
\beq
\int_{\Sigma}(\Psi_1)_{\Omega}\mbox{}^*(\Psi_2)_{\Omega}\Omega^{-2}\sqrt{{\cal M}_{\Omega}}d^{k}x \mbox{ } .  
\label{5}
\eeq  
This inner product additionally succeeds in being banal conformally invariant, being equal to 
\beq
\int_{\Sigma}\Psi_{1}\mbox{}^*\Omega^{\frac{2 - k}{2}} \Psi_{2}\Omega^{\frac{2 - k}{2}}
             \Omega^{-2}\sqrt{{\cal M}}\Omega^{k}\d^{k}x  = 
\int_{\Sigma}\Psi_{1}\mbox{}^*\Psi_{2}\sqrt{{\cal M}}\d^{k}x \mbox{ } 
\eeq
in the `original or mechanically natural' representation in which $\fE$ comes with the trivial 
weight function, 1.
I should caution however that, while this conformal choice of ordering is a nice choice which I am 
suggesting lies on a more solid principle than how it is usually presented (fully explained in 
\cite{Banal}), it might nevertheless eventually be found to clash with other requirements such as 
existence and suitable behaviour of crucial operators.

\noindent
C) Next, and talking only in the context of finite models such as in this paper rather than for full 
geometrodynamics for which the mathematical machinery is lacking,\footnote{The Loop Quantum Gravity 
approach \cite{Thiemann} is advantageous at this point in being equipped with a Hilbert space structure, 
as well as being better-suited as regards A) above.   
Its passage to Ashtekar variables renders operator-ordering issues there different to those of 
geometrodynamics and the toy model of this paper.
Nevertheless, discussion of analogies with the geometrodynamical Wheeler-DeWitt equation as regards the 
Problem of Time remain reasonable and commonplace in the literature, e.g. \cite{K92, Rovellibook, 
Kieferbook} make use of such.}
in general $\widehat{\fH}_{\Omega}$ 
is not self-adjoint with respect to $\mbox{}_{\Omega}\langle \mbox{ } | \mbox{ } \rangle_{\Omega}$, 
while the mechanically-natural $\widehat{\fH}$ is, in a simple sense, with respect to $\langle 
\mbox{ } | \mbox{ } \rangle$.  
I.e. in the sense that  $\int\sqrt{{\cal M}}\d^kx \Psi^*D^2\Psi = 
\int\sqrt{{\cal M}}\d^kx \{D^2\Psi^*\}^2\Psi$ + boundary terms, which amounts to self-adjointness if the 
boundary terms can be arranged to be zero (which is definitely not a problem in this paper as there are 
no boundaries) and in other cases involves such as suitable fall-off conditions on $\Psi$.   
This is not shared by the $\Omega$-inner product as that has an extra factor of $\Omega^{-2}$, which in 
general interferes with the corresponding move by the product rule ($\sqrt{{\cal M}}$ does not interfere 
thus above, since the Laplacian is built out of derivatives that are covariant with respect to the 
metric ${\cal M}_{\sfA\sfB}$.)
However, solving $\fH_{\Omega}\Psi_{\Omega} = \fE_{\Omega}\Psi_{\Omega}$ is equivalent to solving 
$\fH\Psi = \fE\Psi$, so the banal conformal transformation might at this level be viewed as a 
sometimes-useful computational aid, with the answer then being placed in the mechanically-natural 
representation for further physical interpretation.  
This is not an issue in this paper as $\Omega$ is but $1/4$ in my spherical calculations, thus not 
presenting any product-rule obstacles to self-adjointness.    


Now, returning to the Problem of Time, while many conceptual strategies have been put forward to 
resolve it, there is a long history of proposed resolutions not standing up to detailed examination 
\cite{K92}, so that this remains an open problem for GR.  
Some of these strategies are as follows.      

\noindent 1) Perhaps within GR at the classical level there is a {\it fundamental hidden time}.    
E.g., one could seek for such a time by canonically transforming the geometrodynamical variables 
to new variables among which an explicit and genuinely time-like time variable is isolated out.  
One candidate time of this form is the {\it York time}, \cite{York72, K81, K92}.      
This is proportional to $h_{\mu\nu}\pi^{\mu\nu}/{\sqrt{h}}$ so it is a `dilational object'. 
%

\noindent 
2) Perhaps instead there is no fundamental time in Quantum GR but a notion of time {\sl emerges} in the 
quantum regime e.g. in regions of the universe that behave semiclassically \cite{Kieferbook, HallHaw}. 
In situations in which the Born--Oppenheimer approximation 
$\Psi = \psi(\mbox{H}_{\sfA^{\prime}})|\chi(\mbox{H}_{\sfA^{\prime}}, \mbox{L}_{\sfA^{\prime\prime}})
\rangle$ for $\mbox{H}_{\sfA^{\prime}}$ `heavy, slow' and $\mbox{L}_{\sfA^{\prime\prime}}$ `light and 
fast' degrees of freedom and the WKB approximation $\psi(\mbox{H}_{\sfA^{\prime}}) = 
\mbox{e}^{i\sfF(\mbox{\scriptsize H}_{\tfA^{\prime}})/\hbar}$ are applicable, an emergent time drops out 
of the WDE \cite{DeWitt, HallHaw, Kieferbook, SemiclI}.  
For (e.g. using $h_{\mu\nu}$ as `H' and the matter as `L', and requiring that $|\chi\rangle$ depends 
nontrivially on H so that the QM is rendered nonseparable in H, L variables),
\beq
\hbar^2{\cal N}^{\mu\nu\rho\sigma}\frac{\delta^2\Psi}{\delta h^{\mu\nu}\delta h^{\rho\sigma}}
\mbox{ } \mbox{  contains } \mbox{ } 
\hbar^2{\cal N}^{\mu\nu\rho\sigma} \frac{i}{\hbar}\frac{\delta \fF}{\delta h^{\mu\nu}}
\frac{\delta|\chi\rangle}{\delta h^{\rho\sigma}} = 
i\hbar {\cal N}^{\mu\nu\rho\sigma}\pi_{\mu\nu}\frac{\delta|\chi\rangle}{\delta h^{\rho\sigma}}
\label{CruCro}
\eeq 
(by identifying $\fF \approx \fW/M$ where $\fW = \fW(\mH_{\sfA^{\prime}})$ is Hamilton's principal function and 
$M$ is a generic H-mass, and then using the Hamilton--Jacobi relation for the momentum and 
\beq
\left.
\frac{\delta }{\delta t^{\mbox{\tiny WKB}}(h_{\mu\nu})} = 
\frac{1}{\dot{I}}\frac{\pa}{\pa\lambda}
\right) \mbox{ } .  
\eeq 
Then by the momentum--velocity relation (I.10), this expression contains 
\beq 
i\hbar \frac{\delta h^{\rho\sigma}}{\delta t^{\mbox{\tiny WKB}}}
\frac{\delta|\chi\rangle}{\delta h^{\rho\sigma}} = 
i\hbar\frac{\delta|\chi\rangle}{\delta t^{\mbox{\tiny WKB}}}
\eeq 
by the chain-rule in reverse so that one has a TDSE for the local L degrees of freedom with 
respect to a time standard that is (approximately) provided by the background H degrees of freedom.  
An issue here is that (semi)classical conditions need not always occur -- guarantee of a classical 
`large' as in the Copenhagen Interpretation of QM has been cast aside in considering the universe as a 
whole, and will then by no means be recovered in all possible situations.   
The above {\it semiclassical approach} is additionally a useful framework for discussing the origin 
\cite{HallHaw} of galaxies and cosmic microwave background perturbations within the semiclassical 
scheme, for which one needs to study spatially-located fast light degrees of freedom that are coupled 
to global slow heavy degees of freedom such as the size of the universe.   

\noindent 3) There are also timeless records strategies \cite{PW83, GMH, B94I, B94II, EOT, H99, Records}.  
Here, the primary objects are {\it records} -- information-containing subconfigurations of a single 
instant that are localized in both space and configuration space.   
One would then seek to construct a semblance of dynamics or history from the correlations between such 
records.  
QM probability density functions on shape space are here of interest as regards whether one can 
substantiate Barbour's conjectures \cite{B94II, EOT} about a present populated with time-capsules. 

\noindent 4) There are also approaches in which it is the histories that are 
primary \cite{GMH, Hartle}.
 
\noindent The currently intended applications of this paper's toy models are to approaches 2), 3), 4) 
and combinations thereof.  
However, for completeness and due to current interest, I also outline the following additional family 
of strategies.  

\noindent 5) Distinct timeless approach involve {\it evolving constants of the motion} (a Heisenberg 
rather than Schr\"{o}dinger type approach) or {\it partial observables} \cite{Rovellibook}, which is 
used in Loop Quantum Gravity's {\it Master Constraint program} \cite{Thiemann}.  


Quantum GR being technically difficult, toy models such as that being developed in the current paper 
have been useful toward developing strategies such as the above.
Relational particle mechanics are useful such, due to geometrodynamics being parallely formulable 
in relational terms (see \cite{RWR, ABF(K)O} and \cite{RWR, ABVanLan2} for its robustness to the 
inclusion of matter, or for a summary, Sec I.2.3--4).  
Specific Problem of Time applications and extensions of this analogy then feature in \cite{BS89, K92, 
B94I, B94II, EOT, Paris, 06I, 06II, SemiclI, SemiclII, MGM, SemiclIII, Records}). 
In particular, both relational particle mechanics and geometrodynamics have a constraint that depends 
quadratically but not linearly on the momenta, which feature underlies the frozen formalism aspect of 
the Problem of Time, and both have further nontrivial constraints that are linear in the momenta, which 
cause many of the complications with strategies proposed to resolve the Problem of Time.  
Relational particle mechanics have so far been useful toy models in possessing an analogue of the 
above-mentioned dilational York time \cite{06II, SemiclI, SemiclII}, for the semiclassical emergent time 
approach \cite{SemiclI, MGM, SemiclIII} and for the timeless records theory approach \cite{B94I, B94II, 
EOT, Records}.   
See the Conclusions of this paper and \cite{08III} for brief discussion of useful applications of 
(extensions of) these papers' models to these issues.  
In relational particle mechanics models, by their constraints and the subsequent appearance of more 
complicated reduced configuration spaces on which these models' partial observables would live, the 
partial observables approach to these examples of nonrelativistic mechanics would be rather distinct 
from that for the mechanics models considered in \cite{Rovellibook}, and may lead to a broader view than 
therein on what features distinguish nonrelativistic mechanics theories from relativistic ones.


Both the absolute versus relative motion debate and the study of conceptual strategies suggested toward 
resolving the Problem of Time in Quantum Gravity benefit from study of explicit examples of quantum 
relational particle mechanics.  
I provide such in this paper (similarity relational particle mechanics) and \cite{08III} (Euclidean 
relational particle mechanics), by applying my recent understanding of reduced configuration spaces 
\cite{FORD} to carry out quantization. 
[The above-mentioned QM investigations of relational particle mechanics \cite{Smolin, Rovelli, BS89, 06I}, 
and the recent preprint \cite{Gryb} (path integral method), carry out their quantization by other means.]
In this paper I provide Schr\"{o}dinger equations for scalefree {\it N-stop metroland} (relational 
mechanics of N particles in 1-d)  and scalefree {\it N-a-gonland} (relational mechanics of N particles 
in 2-d), concentrating on the latter's triangleland case (3 particles in 2-d).  
For this I discuss how relative angle independent potentials are a substantial simplification (in close 
analogy with how central potentials are in ordinary QM), and provide some simple solutions.  
I begin with relative angle independent multiple harmonic oscillator type potentials, for which I additionally use  
asymptotic, perturbative and numerical methods.  
It is important for this work that I identified my problem to have the same mathematics as the 
Stark effect for the linear rigid rotor.  
In Sec 3 I use rotated coordinates/normal modes/adapted bases to remove the relative angle 
independence restriction: any multiple harmonic oscillator for scalefree triangleland admits coordinates in which it 
takes the form of Sec 2's special case, which amounts to the `electric field' being in an arbitrary 
direction rather than in the simplifying adapted basis in which it points along the Z-axis (of the 
$\mathbb{R}^3$ embedding of the configuration space sphere).  
This is of importance as regards setting up models of the semiclassical approach and records theory.  
My Conclusion (Sec 4) includes an outline of several further examples to which techniques of this paper 
can be applied and which are also useful as regards modelling the Problem of Time and Quantum Cosmology.

\section{Quantum Similarity Relational Particle Mechanics}

\subsection{Time-independent Schr\"{o}dinger equations for Scalefree N-stop metroland and N-a-gonland}

For scalefree N-stop metroland, the configuration space is $\mathbb{S}^{\sn - 1}$ and the conformal-ordered 
time-independent Schr\"{o}dinger equation is (via App A.1 and the notation of Sec I.3.2 and with $\fE$ being 
redefined to absorb the conformal contribution, which is possible as spheres are spaces of constant 
curvature):  
\beq
-\frac{\hbar^2}{2}
\frac{1}{\mbox{sin}^{\sn - 1 - \barr}\Theta_A\prod_{\hat{p} = 1}^{\barr - 1}\mbox{sin}^2\Theta_{\hat{p}}}
\frac{\pa}{\pa\Theta_{\barr}}\left\{\mbox{sin}^{\sn - 1 - \barr}\Theta_{\barr}
\frac{\pa}{\pa\Theta_{\barr}}\right\} + \fV\Psi = \fE\Psi \mbox{ } . 
\label{TISE2}
\eeq
For scalefree N-a-gonland (N $\geq$ 3), the configuration space is $\mathbb{CP}^{\sn - 1}$ and the 
conformal-ordered time-independent Schr\"{o}dinger equation is (via App A.1 and likewise being able to 
absorb the conformal contribution as complex projective spaces are Einstein and hence of constant 
Ricci-scalar curvature \cite{FORD})

\noindent
$$
-\frac{\hbar^2}{2}
\frac{\{1 + ||{\cal R}||^2\}^{2\{\sn - 1\}}}{\prod_{\tip = 1}^{n - 1}{\cal R}_{\barp}}
\left\{
\frac{\pa}{\pa {\cal R}_{\barp}}
\left\{
\frac{\prod_{\barp = 1}^{\sn - 1}{\cal R}_{\barp}}{\{1 + ||{\cal R}||^2\}^{2\sn - 3}}
\{\delta^{\barp\barq} + {\cal R}^{\barp}{\cal R}^{\barq}\}
\frac{\pa\Psi}{\pa{\cal R}_{\barq}}
\right\}
\right. +
$$
\beq
\left.   
\frac{\pa}{\pa {\Theta}_{\tip}}
\left\{
\frac{\prod_{\barp = 1}^{\sn - 1}{\cal R}_{\barp}}{\{1 + ||{\cal R}||^2\}^{2\sn - 3}}
\left\{
\frac{\delta^{\tip\tiq}}{{\cal R}_{\barp}^2} + {1|}^{\tip\tiq}
\right\}
\frac{\pa\Psi}{\pa{\Theta}_{\tiq}}
\right\}  
\right\} + \fV\Psi = \fE\Psi \mbox{ } . 
\label{TISE3}
\eeq

\subsection{Time-independent Schr\"{o}dinger equation for scalefree triangleland}

In this paper we mostly consider the triangleland case for which there is the additional 
`accident' $\mathbb{CP}^1 = \mathbb{S}^2$, which permits use of both spherical type and complex 
projective type variables and carries stronger guarantees of good mathematical behaviour.  
In this case, in spherical coordinates $\{\Theta, \mbox{ } \Phi\}$ and using the `barred 
banal conformal representation' $\overline{\fT} = 4\fT$, $\overline{\fE} + \overline{\fU} = \{\fE + \fU\}/4$, 
the time-independent Schr\"{o}dinger equation is
\beq
- \frac{\hbar^2}{2}\hat{\cal J}^2\Psi = -\frac{\hbar^2}{2
}\left\{\frac{1}{\mbox{sin}\Theta}\frac{\pa}{\pa\Theta}
\left\{
\mbox{sin}\Theta\frac{\pa\Psi}{\pa\Theta}
\right\} 
+ \frac{1}{\mbox{sin}^2\Theta}\frac{\pa^2\Psi}{\pa\Phi^2}  
\right\} = \{\overline{\fE}(\Theta) + \overline{\fU}(\Phi, \Theta)\}\Psi \mbox{ } .  
\label{spheTISE}
\eeq
[Now $\fE$ does {\it not} need redefining since the conformal term is zero because the configuration 
space dimension is 2.]  
While in plane polar coordinates $\{{\cal R}, \Phi \}$ obtained by passing to stereographic coordinates 
on the sphere and then passing to the `tilded banal conformal representation'  
$\widetilde{\fT} = \fT\{1 + {\cal R}^2\}^2$ and 
$\widetilde{\fE} + \widetilde{\fU} = \{\fE + \fU\}/\{1 + {\cal R}^2\}^2$, the time-independent Schr\"{o}dinger equation is   
\beq
- \frac{\hbar^2}{2}\left\{\frac{1}{{\cal R}}\frac{\pa}{\pa{\cal R}}
\left\{
{\cal R}\frac{\pa\Psi}{\pa{\cal R}}
\right\} 
+ \frac{1}{{\cal R}^2}\frac{\pa^2\Psi}{\pa\Phi^2}
\right\} = \{\widetilde{\fE}({\cal R}) + \widetilde{\fU}({\cal R})\}\Psi \mbox{ } .  
\label{flatTISE}
\eeq

\subsection{Inner products for scalefree triangleland in various coordinate systems}

The Jacobians in question are to be read off the metrics in Sec I.3.2. 
In the barred banal conformal representation, this gives in spherical coordinates the well-known 
$\mbox{sin}\Theta$ factor, in stereographic coordinates it gives ${\cal R}/\{1 + {\cal R}^2\}^2$ 
(which combines the usual ${\cal R}$ of plane polar coordinates with the requisite conformal factors), 
and, if one uses $I_1$ (or $I_2$ or $\iota_1$ or $\iota_2$) instead of ${\cal R}$, it gives but a 
constant factor.  
As there is no nontrivial weight in this representation, the above are the entirety of the extra factors 
inside the inner product (up to proportion, which one would then fix by normalization).  
In the tilded banal conformal representation, the Jacobians are the usual ${\cal R}$ in plane polar 
coordinates$({\cal R}, \Phi)$, $\mbox{sec}^4\mbox{$\frac{\Theta}{2}$}\mbox{sin}\Theta$ in spherical 
coordinates (which combines the usual sin$\Theta$ with the requisite conformal factors), if one uses 
$\mI_1$, it gives $1/\{\mI - \mI_1\}^2$ up to proportionality, and, if one uses $\mI_2$, it gives 
$1/\mI_2^2$.  
However this representation also has a nontrivial weight [c.f. (\ref{5})], so that one ends up with 
precisely the same inner product as for the barred banal conformal representation (as should be the 
case, for in configuration space dimension 2, the wavefunctions themselves do not 
scale).

\subsection{Separability for $\Phi$-independent potentials}

Then the above time-independent Schr\"{o}dinger equations are separable under the separation ans\"{a}tze 
\beq
\Psi({\cal R}, \Phi) = \zeta({\cal R})\eta(\Phi) \mbox{ } \mbox{ or } \mbox{ } 
\Psi(\Theta, \Phi) = \xi(\Theta)\eta(\Phi) \mbox{ } ,
\eeq 
and in each case one obtains simple harmonic motion solved by 
\beq
\eta = \mbox{exp}(\pm i\mj\Phi)
\label{SHM}
\eeq 
for j an integer. 
This is a relative angular momentum quantum number corresponding to ${\cal J}$ being classically 
conserved, in analogy with how there is an angular momentum quantum number m in the central-potential 
case of ordinary QM corresponding to the angular momentum $\mL_z$ being classically conserved.   
The accompanying separated-out equation is, in the tilded banal conformal representation in 
(${\cal R}$, $\Phi$) coordinates, the radial equation 
\beq
{\cal R}^2\zeta_{\cal RR} + {\cal R}\zeta_{\cal R} - 
\{
2{\cal R}^2\{\widetilde{\fV}({\cal R}) - \widetilde{\fE}({\cal R})\}/{\hbar^2} + \mj^2
\}
\zeta = 0 \mbox{ } ,  
\label{squib}
\eeq
or, in the barred conformal representation in ($\Theta$, $\Phi$) coordinates, the azimuthal equation 
\beq
\{\mbox{sin}\Theta\}^{-1}\{\mbox{sin}\Theta\xi_{\Theta}\}_{\Theta} - 
\{2\widetilde{\fV}(\Theta) - \fE\}/\hbar^2 + \mj^2\{\mbox{sin}\Theta\}^{-2}\}\xi = 0  \mbox{ } .  
\eeq

\noindent{\large\bf 2.5 The special harmonic oscillator quantum problem}

\mbox{ }

\noindent
The problem is, in the barred banal conformal representation in spherical coordinates, (\ref{spheTISE}) 
with an harmonic oscillator type potential (I.59) inserted in, which I rearrange to the dimensionless form 
\beq
\frac{1}{\mbox{sin}\Theta}\frac{\pa}{\pa\Theta}
\left\{ 
\mbox{sin}\Theta\frac{\pa\Psi}{\pa\Theta}  
\right\} 
+ \frac{1}{\mbox{sin}^2\Theta}\frac{\pa^2\Psi}{\pa\Phi^2} 
+ \{{\cal E} - {\cal A} - {\cal B}\mbox{cos}\Theta\}\Psi = 0
\label{B}
\eeq

\noindent
for 
\beq
{\cal A} \equiv 2A/\hbar^2 \mbox{ } , \mbox{ } 
{\cal B} \equiv 2B/\hbar^2 \mbox{ } ,  \mbox{ } 
\overline{{\cal E}} \equiv 2\overline{\fE}/\hbar^2 \mbox{ } .  
\label{ABE}
\eeq
This then separates into (\ref{SHM}) and
\beq
\{\mbox{sin}\Theta\}^{-1}\{\mbox{sin}\Theta\xi_{\Theta}\}_{\Theta} - 
\{{\cal A} - \overline{{\cal E}} + {\cal B}\mbox{cos}(\Theta) + \mj^2\{\mbox{sin}\Theta\}^{-2}\}\xi = 0 
\mbox{ } .  
\label{AA}
\eeq

Alternatively, in the tilded banal conformal representation in (${\cal R}, \Phi$) coordinates, 
it is (\ref{flatTISE}) with (I.58) inserted in, which I rearrange to the dimensionless form 
\beq
\frac{1}{{\cal R}}\frac{\pa}{\pa{\cal R}}
\left\{
{\cal R}\frac{\pa\Psi}{\pa{\cal R}}
\right\}
+ \frac{1}{{\cal R}^2}\frac{\pa^2\Psi}{\pa\Phi^2} - 
\left\{ 
\frac{{\cal K}_1 + {\cal K}_2{\cal R}^2}{\{1 + {\cal R}^2\}^2} - 
\frac{{\cal E}}{\{1 + {\cal R}^2\}^2} 
\right\}\Psi = 0 \mbox{ } , 
\eeq
for 
\beq
{\cal K}_i = K_i/\hbar^2 \mbox{ } \mbox{  and } \mbox{ } {\cal E} = 2\fE/\hbar^2 \mbox{ } .  
\eeq
This then separates into (\ref{SHM}) and
\beq
{\cal R}^2\zeta_{\cal RR} + {\cal R}\zeta_{\cal R} - 
\left\{
\frac{{\cal K}_1 + {\cal K}_2{\cal R}^2}{\{1 + {\cal R}^2\}^2} - 
\frac{\widetilde{\fE}}{\{1 + {\cal R}^2\}^2} + \mj^2
\right\}
\zeta = 0 \mbox{ } .  
\label{BB}
\eeq

Note that these equations are self-dual in the sense of \cite{08I}, so that, again, direct study of only 
one of the two asymptotic regimes is necessary and then the other can be read off by mere substitution.    

\mbox{ }

\noindent{\large\bf 2.6 Useful mathematical analogue}

\mbox{ }

\noindent
It is next convenient to recollect my observation \cite{08I} that scalefree triangleland's special triple 
harmonic oscillator like potential has the same mathematical form as the fairly well-known problem of the linear rigid 
rotor in a background homogeneous electric field in the symmetry-adapted basis for that problem.
This then at the quantum level amounts to the present mathematics being analogous to that of the Stark 
effect for the linear rigid rotor, which is well-documented (see e.g. \cite{TSHecht, Messiah} and Secs 
2.5, 2.7, 2.8, 3).  

\mbox{ }

\noindent{\large\bf 2.7 Exact solution for the `very special multiple harmonic oscillator'}

\mbox{ }

\noindent
In this case the analogy is with the linear rigid rotor itself \cite{AF, LLQM}, and 
the spherical representation's mathematics takes a familiar form, (\ref{AA}) now being    
the associated Legendre equation (\ref{Leg}).

One then has a quantum number $\mJ \in \mathbb{N}_0$ analogous to the total angular momentum quantum number l of 
the linear rigid rotor (the analogy is that $\mJ$ {\sl is} a total angular momentum in a 3-d space, but that 3-d space 
is not ordinary 3-d space but rather the scaled triangleland configuration space, c.f. App. I.C).  
This obeys 
\beq
\mJ\{\mJ + 1\} = {\cal E} - {\cal A} = 2\{\overline{\fE} - A\}/\hbar^2
\eeq
and hence 
\beq
\overline{\fE} = {\hbar^2}\mJ\{\mJ + 1\}/2 + A
\eeq
in the spherical representation, or 
\beq
\fE = 2\hbar^2\mJ\{\mJ + 1\} + \{K_1 + K_2\}/{4} \label{mabel}
\eeq
in the planar representation.
One also has a quantum number $\mj \in \mathbb{Z}$ such that $|\mj| \leq \mJ$; this is analogous to the 
magnetic quantum number of the linear rigid rotor, and here has the interpretation of being the 2-d 
relative angular momentum (which is the Z-component of the above `total angular momentum').

Moreover, if $\fE$, $K_1$, $K_2$ are to take the interpretation of being fixed, then there will 
either be 1 or 0 such $\mJ$ -- a closed universe type truncation of the number of allowed states.  
From (\ref{mabel}) then, for there to be any chance of solutions one needs $\fE \geq \{K_1 + K_2\}/4$, 
so for harmonic oscillator type models, $\fE > 0$ is indispensable. 
If there is a J and it is not zero, there are various degenerate solutions corresponding to different 
values of j.
These correspond to states of different relative angular momentum between the particle 2, 3 subsystem 
and the particle 1 subsystem.

The wavefunctions are of the form
\beq
\Psi(\Theta, \Phi) \propto P_{\sJ}^{\sj}(\mbox{cos}\Theta)\mbox{exp}(i\mbox{j}\Phi)
\label{pummel} \mbox{ } . 
\eeq
Thus, in the planar representation, the solution is  
\beq
\Psi_{\sJ\sj}({\cal R}, \Phi) \propto  P_{\sJ}^{\sj}
\left(
\frac{1 - {\cal R}^2}{1 + {\cal R}^2}
\right)
\mbox{exp}(i\mbox{j}\Phi) \mbox{ } , 
\eeq
and, in terms of the physically and visually useful partial moments of inertia--relative angle 
variables, it is 

\noindent
\beq
\Psi_{\sJ\sj}(\iota_1, \iota_2, \Phi) \propto  
P_{\sJ}^{\sj}
\left(
\frac{\mI_2 - \mI_1}{\mI_1 + \mI_2}
\right)
\mbox{exp}(i\mj\Phi) =
P_{\sJ}^{\sj}
\left(
\frac{\mI - 2\mI_1}{\mI}
\right)
\mbox{exp}(i\mj\Phi) = 
P_{\sJ}^{\sj}
\left(
\frac{2\mI_2 - \mI}{\mI}
\right)
\mbox{exp}(i\mj\Phi) \mbox{ } .  
\eeq
While, in terms of the original variables of the problem,  
\beq
\Psi_{\sll\sj}(\underline{\iota}_1, \underline{\iota}_2) \propto  
P_{\sJ}^{\sj}
\left(
\frac{||\underline{\iota}_2||^2 - ||\underline{\iota}_1||^2}
     {||\underline{\iota}_1||^2 + ||\underline{\iota}_2||^2}
\right)
\mbox{exp}
\left(
i\mj\mbox{ arccos}
\left(
 \frac{\underline{\iota}_1\cdot\underline{\iota}_2}{||\underline{\iota}_1||||\underline{\iota}_2||}
\right)
\right) \mbox{ } .  
\eeq

\mbox{ }

\noindent{\footnotesize[{\bf Figure 1}. For the first three values of $\mJ$ (0, 1 and 2), I provide 

\noindent a) the azimuthal probability density function on the sphere, which is standard.

\noindent b) The radial probability density function in the $({\cal R}, \Phi)$ plane
Unlike the usual situation with the radial probability density function in the atom, the inner peaks are 
the taller ones (in the atom it's more probable e.g. that the 2s electron is `outside' the 1s one).  
This is due to the unusual inner product of our planar problem.  

\noindent c) 
These plots can then be interpreted in terms of straightforward relational variables such as 
$\iota_1, \iota_2$, $\mI_1$ or $\mI_2$.  
I plot in terms of $\mI_1$.
Note the reflection symmetry about I/2 and also that plots in terms of $\mI_2$ coincide with those 
in terms of $\mI_1$ and so I do not provide them.   
From the various above plots, one can infer on which of the sorts of triangles defined in Fig I.5d) 
the various wavefunctions peak.
The cases depicted exhibit one of the following patterns of likelihood for configurations (triangles): 
a Jacobi-regular peak (if $|\mj|$ = J), 
a Jacobi-tall and a Jacobi-flat peak with an interposed Jacobi-regular node (if $|\mj|$ = J -- 1), 
a Jacobi-tall peak, a Jacobi-regular peak and a Jacobi-flat peak with two interposed nodes 
of more moderate tallness and flatness ($|\mj|$ = J -- 2).   
Note that the higher J is, the more pronounced the tallness and flatness involved is.  
Next, note that for j = 0 all the wavefunctions are surfaces of revolution.
Higher values of j have sinusoidal dependence on $\Phi$.  
Moreover the directions picked out by this have physical meaning.  
$|\mj|$ = 1 has what would usually be $p_x$' and $p_y$ (or $d_{xz}$ and $d_{yz}$) directionality in space, 
which in our problem signifies near-collinear peaking and near-isosceles peaking in the 
configuration space of triangles. 
While, $|\mj|$ = 2 picks out both of the above equally ($d_{xy}$) or avoids both equally 
($d_{x^2 - y^2}$).]}  

\mbox{ } 

\noindent Note 1) paralleling the classical working, the very special solution is unconditionally self-dual 
under the duality map.

\noindent 2) while this case is simple and not general, its exact solution serves as something about 
which one can conduct a more general perturbative treatment (SSec 2.10).  

\mbox{ }

\noindent{\bf\large 2.8 Small asymptotics solutions for the `special multiple harmonic oscillator'}

\mbox{ } 

\noindent 
I work in the tilded $\fQ$-representation. 
In the first small approximation, $2\{\widetilde{\fE} + \widetilde{\fU}\} = Q_0$ (a constant evaluated  
in Sec I.4.9). 
Thus, in terms of ${\cal Q}_0 \equiv Q_0/\hbar^2$, the time-independent Schr\"{o}dinger equation is 
\beq
\frac{1}{{\cal R}} \frac{\pa}{\pa{\cal R}} 
\left\{
{\cal R}\frac{\pa\Psi}{\pa{\cal R}}
\right\} 
+ \frac{1}{{\cal R}^2}\frac{\pa^2\Psi}{\pa\Phi^2} + {\cal Q}_0\Psi = 0
\mbox{ } ,
\eeq
which is a familiar problem, separating into simple harmonic motion (\ref{SHM}) and a radial equation (\ref{squib}), which 
is now the Bessel equation of order j (\ref{Bess}) in the rescaled variable  
${\cal S} = \sqrt{{\cal Q}_0}{\cal R}$. 
Thus there is just the one quantum number $\mj \in \mathbb{Z}$ (relative angular momentum).  
The wavefunctions are then 
\beq
\Psi_{\sj}({\cal R}, \Phi) \propto J_{\sj}(\sqrt{{\cal Q}_0}{\cal R})\mbox{exp}(i\mj\Phi) 
\mbox{ } .
\eeq
So, in spherical variables, 
\beq
\Psi_{\sj}(\Theta, \Phi) \propto J_{\sj}(\sqrt{{\cal Q}_0} \mbox{tan$\frac{\Theta}{2}$} )
\mbox{exp}(i\mj\Phi) \mbox{ } ,
\eeq
and in terms of the partial moments of inertia--relative angle variables,
it is 
\beq
\Psi_{\sj}(\mI_1, \mI_2, \Phi) \propto  
J_{\sj}(\sqrt{{\cal Q}_0\mI_1/\mI_2})\mbox{exp}(i\mj\Phi) = 
J_{\sj}(\sqrt{{\cal Q}_0\mI_1/\{\mI - \mI_1\}})\mbox{exp}(i\mj\Phi) =
J_{\sj}(\sqrt{{\cal Q}_0\{\mI - \mI_2\}/\mI_2})\mbox{exp}(i\mj\Phi)
\mbox{ } . 
\eeq
While, in terms of the original variables of the problem, 
\beq
\Psi_{\sj}(\underline{\iota}_1, \underline{\iota}_2) \propto  J_{\sj}
\left(
\sqrt{{\cal Q}_0}
\left\|
\frac{\underline{\iota}_1}{\underline{\iota}_2}
\right\|
\right)
\mbox{exp}
\left(
i\mj\mbox{ arccos}
\left(
\frac{\underline{\iota}_1\cdot\underline{\iota}_2}{||\underline{\iota}_1||||\underline{\iota}_2||}
\right)
\right) 
\mbox{ } .  
\label{Mabelle}
\eeq
I sketch the probability density function in various of these variables in Fig 2.  
Finally note that replacing the specific constant $Q_0$ by some other constant $Q$, this also provides the exact solution to 
the constant potential problem.

\mbox{ }  

In the second approximation, $2\{\widetilde{\fE} + \widetilde{\fU}\} = Q_0 - Q_2{\cal R}$, where $Q_2$ 
is another constant evaluated in Sec I.4.9, and which I consider here to be strictly positive.  
In this case the time-independent Schr\"{o}dinger equation is
\beq
\frac{1}{{\cal R}}\frac{\pa}{\pa{\cal R}}\left\{{\cal R}\frac{\pa\Psi}{\pa{\cal R}}\right\} 
+ \frac{1}{{\cal R}^2}\frac{\pa^2\Psi}{\pa\Phi^2} + \{{\cal Q}_0 - {\cal Q}_2{\cal R}^2\}
\Psi = 0 \mbox{ } ,
\eeq
(where I use ${\cal Q}_2 \equiv Q_2/\hbar^2$), which is also a familiar equation, straightforwardly mapping 
to the 2-d isotropic harmonic oscillator \cite{Messiah, Schwinger, Robinett} by the correspondence in App B.  
It separates into simple harmonic motion (\ref{SHM}) and a radial equation (\ref{squib}) that this time can be mapped to 
the associated Laguerre equation (\ref{Laguerre}).  
Thus this problem's solution is as follows. 
In addition to the relative angular momentum quantum number j $\in \mathbb{Z}$, there is a `radial', `principal' or 
`node-counting' quantum number R $\in \mathbb{N}_0$ such that 
\beq
Q_0/2\hbar\sqrt{Q_2} = |\mj| + 2\mR + 1
\label{Arya}
\eeq
holds. 
In terms of the original quantities of the problem, this corresponds to 
\beq
\fE = {K_2}/{2} + \hbar\sqrt{4\fE + K_1 - 3K_2}\{|\mj| + 2\mR + 1\}
\eeq
or, in the language of the spherical problem
\beq
\overline{\fE} = A + B + \hbar\sqrt{\overline{\fE} - A - 2B}\{|\mj| + 2\mR + 1\} \mbox{ } .  
\eeq
Then  
\beq
\fE \geq {K_2}/{2} + \hbar\sqrt{4\fE + K_1 - 3K_2} \mbox{ } ( \mbox{ } > 0 \mbox{ } )
\eeq
is necessary for there to be any solutions at all.  
While, the non-standard interpretation that all of $\fE, K_1, K_2$ are fixed will restrict the number 
of simultaneously-relevant solutions in my closed-universe context; e.g. rational--irrational 
incompatibility can be used to construct cases with no solutions, but there are also cases for which 
there remain interesting degenerate possibilities such as (R, j) = (0, $\pm$ 2) and (1, 0).

The wavefunctions are
\beq
\Psi_{\sR\sj}({\cal R}, \Phi) \propto 
{\cal R}^{|\sj|}\mbox{exp}(\sqrt{{\cal Q}_2}{\cal R}^2/2)L_{\sR}^{|\sj|}(\sqrt{{\cal Q}_2 }{\cal R}^2)
\mbox{exp}(i\mj\Phi)  
\eeq
or, in spherical variables, 
\beq
\Psi_{\sR\sj}(\Theta, \Phi) \propto
\mbox{tan}^{|\sj|}\mbox{$\frac{\Theta}{2}$}\mbox{exp}({\cal Q}_2\mbox{tan}^2\mbox{$\frac{\Theta}{2}$}/2)
L^{|\sj|}_{\sR}(\sqrt{{\cal Q}_2}\mbox{tan}^2\mbox{$\frac{\Theta}{2}$})\mbox{exp}(i\mj\Phi) 
\mbox{ } ,
\eeq
while in terms of the partial moments of inertia--relative angle variables,
it is 
$$
\Psi_{\sr\sj}(\mI_1, \mI_2, \Phi) \propto  
\left\{
\frac{\mI_1}{\mI_2}
\right\}^{|\sj|/2}
\mbox{exp}
\left(
-\frac{\sqrt{{\cal Q}_2}}{2}
\frac{\mI_1}{\mI_2}
\right)
L^{|\sj|}_{\sR}
\left(
\sqrt{ {\cal Q}_2}
\frac{\mI_1}{\mI_2}
\right)
\mbox{exp}(i\mj\Phi) =
$$
\beq
\left\{
\frac{\mI_1}{\mI - \mI_1}
\right\}^{|\sj|/2}
\mbox{exp}
\left(
-\frac{\sqrt{{\cal Q}_2}}{2}
\frac{\mI_1}{\mI - \mI_1}
\right)
L^{|\sj|}_{\sR}
\left(
\sqrt{ {\cal Q}_2}
\frac{\mI_1}{\mI - \mI_1}
\right)
\mbox{exp}(i\mj\Phi) = 
\left\{
\frac{\mI - \mI_2}{\mI_2}
\right\}^{|\sj|/2}
\mbox{exp}
\left(
-\frac{\sqrt{{\cal Q}_2}}{2}
\frac{\mI - \mI_2}{\mI_2}
\right)
L^{|\sj|}_{\sR}
\left(
\sqrt{ {\cal Q}_2}
\frac{\mI - \mI_2}{\mI_2}
\right)
\mbox{exp}(i\mj\Phi)  \mbox{ } .  
\eeq
While, in terms of the original coordinates for the problem,
\beq
\Psi_{\sr\sj}(\underline{\iota}_1, \underline{\iota}_2) \propto 
\left\|
\frac{\underline{\iota}_1}{\underline{\iota}_2}
\right\|^{|\sj|}
\mbox{exp}
\left(
-\frac{\sqrt{{\cal Q}_2}  }{2}
\left\|\frac{\underline{\iota}_1^2}{\underline{\iota}_2^2}\right\|
\right)
L^{|\sj|}_{\sR}
\left(
\frac{\sqrt{{\cal Q}_2}  }{\hbar}
\left\|\frac{\underline{\iota}_1^2}{\underline{\iota}_2^2}\right\|
\right)
\mbox{exp}
\left(
i\mbox{j}\mbox{ arccos}
\left(
\frac{
\underline{\iota}_1\cdot\underline{\iota}_2}{||\underline{\iota_1}||||\underline{\iota}_2||}
\right) 
\right) \mbox{ } .  
\eeq

That a 2-d isotropic harmonic oscillator with eigenvalues (\ref{Arya}) resides within the scalefree triangleland multiple 
harmonic oscillator like problem as a limit problem has counterpart in the literature on the linear rotor   
(\cite{PS57}, see also the figure in \cite{Meyenn}).

\mbox{ }

\noindent{\footnotesize[{\bf Figure 2} a) Sketches of probability density functions for the first approximate solutions in terms 
of ${\cal R}$ for j = 0 and j = 1 (which is qualitatively similar to j = 2).
For ${\cal Q}_0 = 1$, what the relevant region contains is but the first bit of ascension to the 
first of an infinity of peaks (first two subfigures).   
For ${\cal Q}_0$ large various whole peaks can lie in the relevant regime; the two plots I provide 
are for ${\cal Q}_0 = 10^4$ with j = 0 and j = 1.    

\noindent b) Sketch of the (unnormalized) probability density functions for the first approximate solutions in terms of $\mI_1$ for j = 0 and 
j = 1, for ${\cal Q}_0 = 1$ and $10^4$.  
Note that for the first approximate solution in terms of $\mI_2$, all one need do is reflect in I/2 
(and the relevant region maps to the new relevant region and so is the same portion as before).  
Thus I do not provide separate figures for these but rather provide the whole range of values of $\mI_1$.  
This reflection trick works throughout this Section.  

\noindent c) The second approximate solutions in terms of ${\cal R}$ for (R, j) = (0, 0), (0, $\pm1$) or 
(0, $\pm2$) and (1, 0) for ${\cal Q}_2 = 1$ and then for ${\cal Q}_2 = 100$ in $\hbar = 1$ units.

\noindent d) The second approximate solution in terms of $\mI_1$ for these same six cases.]}    

\mbox{ }

\noindent
Note 1) As regards which inner product to use, in the first approximation, one can use the na\"{\i}ve 
${\cal R}$ in place of ${\cal R}/\{1 + {\cal R}^2\}^2$ at the cost of an error of up to 2$\%$ over 
${\cal R} \in [0, 0.1]$.  
This however swamps the improvement in passing from first to second approximation; one should use at 
least ${\cal R}\{1 - 2{\cal R}^2\}$ in the latter case so as to not compromise accuracy.  

\noindent 2) The unusual inner product in use here does cause these probability density function differ 
somewhat from the usual ones for a 2-d constant potential or 2-d isotropic harmonic oscillator: the peaks are somewhat 
shifted inwards, and the inner peaks gain in relative importance; these are signs of greater confinement.  

\noindent 3) How can few-peak functions be second approximations when infinite-peak functions are first 
approximations? 
The trick is that the few-peak functions have an extra node-counting quantum number R, so that the first 
approximation can be seen as a superposition of many different values of R each with its own 
${\cal Q}_2$ as dictated by (\ref{Arya}), and thus build up a multiplicity of peaks.  
For the values considered, there is better than 1$\%$ accuracy between the 2 approximations 
up to some value of ${\cal R}$ of order of magnitude 0.01 to 0.1.  

\mbox{ }

\noindent{\large\bf 2.9 Large asymptotics solutions for special harmonic oscillator and other problems}

\mbox{ }

\noindent
I now apply duality to write down first and second large approximation solutions.  
The first approximation gives 
$$
\Psi_{\sj}({\cal R}, \Phi) \propto  J_{\sj}(\sqrt{{\cal Q}_4}/{\cal R})\mbox{exp}(i\mj\Phi) 
\mbox{ }  \mbox{ or } \mbox{ } 
\Psi_{\sj}(\Theta, \Phi) \propto  J_{\sj}(\sqrt{{\cal Q}_4}\mbox{cot}\mbox{$\frac{\Theta}{2}$})
\mbox{exp}(i\mj\Phi) \mbox{ } . 
$$
Thus, in terms of the physically and visually useful partial moments of inertia--relative angle variables,
it is 
$$
\Psi_{\sj}(\mI_1, \mI_2, \Phi) = J_{\sj}(\sqrt{{\cal Q}_4\mI_2/\mI_1})\mbox{exp}(i\mj\Phi) =
J_{\sj}(\sqrt{{\cal Q}_4\{\mI - \mI_1\}/\mI_1})\mbox{exp}(i\mj\Phi) = 
J_{\sj}(\sqrt{{\cal Q}_4\mI_2/\{\mI - \mI_2\}})\mbox{exp}(i\mj\Phi) 
\mbox{ } . 
$$
While, in terms of the original coordinates of the problem, 
\beq
\Psi_{\sj}(\underline{\iota}_1, \underline{\iota}_2) \propto  J_{\sj}
\left(
\sqrt{{\cal Q}_4}
\left\|
\frac{\underline{\iota_2}}{\underline{\iota_1}}
\right\|
\right)
\mbox{exp}
\left(
i\mj\mbox{ arccos}
\left(
\frac{\underline{\iota_1}\cdot\underline{\iota_2}}{||\underline{\iota_1}||||\underline{\iota_2}||}
\right)
\right) 
\mbox{ } .  
\label{Mabel}
\eeq
\noindent{\footnotesize[{\bf Figure 3} Using the same `radial' inner product as in figure 1, 
I obtain the following probability density functions in terms of ${\cal R}$. [In terms of $\mI_1$, these have the 
same form as the corresponding small solution's probability density function in terms of $\mI_2$ and vice versa, so I do 
not provide any new probability density functions as functions of $\mI_1$, $\mI_2$.]

\noindent a) For the first approximation, I provide an archetypal probability density function for ${\cal Q}_4$ = 1 
and $10^4$.  

\noindent b) For the second approximation, I provide the probability density function's for (R, j) = (0, 0), (0, 1) and (1, 0) 
[(0, 2) is qualitatively similar to (0, 1)] for ${\cal Q}_6 = 1$ and then $10^4$.}]

\mbox{ }  

As happens classically, the above analysis also holds for $\widetilde{\fV}_{(\alpha, 0)}$ with constant 
of proportionality $\Lambda_{\alpha}$, one gets exactly the same large-asymptotics analysis as here, 
with $q_0 = 2\fE - \Lambda_{\alpha}$, $q_2 = 4\fE - \{4 + \alpha\}\Lambda_{\alpha}$.  
Thus the conditional duality map has put the curious large asymptotic region in firmly understood terms.  

\mbox{ }

\noindent{\large\bf 2.10 Special triple harmonic oscillator problem treated perturbatively for small B} 

\mbox{ }

\noindent
Let us treat the special triple harmonic oscillator problem as a perturbation about the simple `very special triple harmonic oscillator'.  
For a perturbation ${\cal H}^{\prime}$ to one's rescaled Hamiltonian [rescaled as the other calligraphic 
quantities in (\ref{ABE}) are] the first few objects of perturbation theory as used in this paper are as 
follows \cite{LLQM}. 
For $\overline{\cal E}_{\sJ} = \overline{\cal E}_{\sJ}^{(0)}$ nondegenerate,  
$
\overline{\cal E}_{\sJ}^{(1)} = \langle \mJ| {\cal H}^{\prime} |\mJ\rangle$, 
$
|\Psi_{\sJ}^{(1)}\rangle = 
- \sum_{\sK \neq \sJ} \frac{    |\sK\rangle \langle \sK| {\cal H}^{\prime} |\sJ\rangle    }
                           {    \overline{\cal E}_{\tK} - \overline{\cal E}_{\tJ}   }
$
at first order and 
$
\overline{\cal E}_{\sJ}^{(2)} = -\sum_{\sK \neq \sJ} \frac{ |\langle \sK| {\cal H}^{\prime} |\sJ\rangle|^2}
                                             {\overline{\cal E}_{\tK} - \overline{\cal E}_{\tJ}} \mbox{ }   
$
at second order.  
For $\overline{\cal E}_{\sJ}^{(0)}$ degenerate, one needs to solve at first order 
$
\sum_{\sss} \langle \mJ, \mj| {\cal H}^{\prime}| \mJ, \ms  \rangle a_{\sss}^{(1)} = 
\overline{\cal E}^{(1)}a_{\sj}^{(1)},
$
and, at second order, 
$
\sum_{\sK \neq \sJ}\sum_{\sss} \frac{      \langle \sK, \sj| {\cal H}^{\prime}| \sJ, \sss  \rangle 
\langle \sJ, \sss| {\cal H}^{\prime}| \sK, \sk  \rangle      }
{      \overline{\cal E}_{\tJ} - \overline{\cal E}_{\tK}      } a_{\tk}^{(2)} = 
\overline{\cal E}^{(2)}a_{\tj}^{(2)} 
$.

For us, the $|\Psi_{\sJ\sj}\rangle = |\mJ, \mj\rangle$ are (\ref{pummel}) and $H^{\prime} = B^{\prime}X$ 
for $X$ the `Legendre variable' related to $\Theta$ by $X = \mbox{cos}\Theta$.  
Then the key integral underlying time-independent perturbation theory is  
$
\langle \Psi_{\sJ^{\prime}\sj^{\prime}} | \fV^{\prime} | \Psi_{\sJ\sj} \rangle 
$
(here there's no subtlety in the style of Sec 2.3 with the inner product as the banal conformal 
transformation here has a merely constant conformal factor and so cancels out upon performing 
normalization).
The above has as its nontrivial factor 
$
\int_{-1}^{1}P_{\sJ^{\prime}}^{\sj^{\prime}}(X)XP_{\sJ}^{\sj}(X)\d X \mbox{ } , 
$
but there is a recurrence relation (\ref{**}) enabling $XP_{\sJ}^{\sj}(X)$ to be turned into a linear 
combination of $P_{\sJ^{\prime\prime}}^{\sj^{\prime\prime}}(X)$, whereupon orthonormality of the 
associated Legendre functions (\ref{*}) can be applied to evaluate it.  
This calculation parallels that in the derivation of selection rules for electric dipole transitions 
\cite{AF, Blum}.  
Then
$
\langle \mJ, \mj^{\prime} |{\cal B}X |\mJ, \mj \rangle = 0 
$
since the recurrence relation sends $XP_{\sJ}^{|\sj|}$ to a sum of $P_{\sJ^{\prime}}^{|\sj^{\prime}|}$ 
for $\mj^{\prime} \neq \mj$, so each contribution to the integral vanishes by orthogonality.  
While, for 
$
\langle \mJ^{\prime}, \mj^{\prime} |BX |\mJ, \mj \rangle
$,   
one similarly needs $\mJ^{\prime} =  \mJ \pm 1$ and $\mj^{\prime} = \mj$ to avoid it vanishing by 
orthogonality.  
So two cases survive this `selection rule'.  
The first is, by direct computation,  
\beq
\langle \mJ + 1, \mj |{\cal B}X |\mJ, \mj \rangle = 
{\cal B}  \sqrt{\frac{\{\mJ + 1\}^2 - \mj^2}{\{2\mJ + 1\}\{2\mJ + 3\}}}
\label{first} \mbox{ } , 
\eeq
while the second is 
\beq
\langle \mJ - 1, \mj |{\cal B}X |\mJ, \mj \rangle = 
\langle \mJ, \mj | {\cal B}X | \mJ - 1, \mj\rangle = 
{\cal B}  \sqrt{\frac{\mJ^2 - \mj^2}{\{2\mJ - 1\}\{2\mJ + 1\}}}  
\eeq
by using (\ref{first}) with J -- 1 in place of J (parallelling e.g. \cite{Messiah}).  

The eigenspectrum is thus
\beq
\overline{\cal E}_{\sJ, \sj} = \mJ\{\mJ + 1\} + {\cal A} + 
\frac{{\cal B}^{2}\{\mJ\{\mJ + 1\} - 3\mj^2\}}{2\mJ\{\mJ + 1\}\{2\mJ - 1\}\{2\mJ + 3\}} + 
O({\cal B}^4)  
\eeq
so 
\beq
\fE_{\sJ, \sj} = 2\hbar^2\mJ\{\mJ + 1\} + 4A + 
\frac{4B^{2}\{\mJ\{\mJ + 1\} - 3\mj^2\}}{\hbar^2\mJ\{\mJ + 1\}\{2\mJ - 1\}\{2\mJ + 3\}} + 
O(B^4)  \mbox{ } .  
\label{Diesel}
\eeq
For $J = 0$, one needs a separate calculation, which gives 
\beq
\fE_{0, 0} =  4A + \frac{4B^{2}}{3\hbar^2} + O(B^4)  \mbox{ } .
\label{Diesel2}
\eeq

%

This calculation can then be checked against its rotor counterpart (originally in \cite{Kronig} 
and which can also found in e.g. \cite{TSHecht, Messiah}].  
The corresponding eigenfunctions can be looked up (e.g. \cite{Proprin}) and reinterpreted in terms of 
the original problem's mechanical variables. 

\mbox{ }

\noindent{\large\bf 2.11 Placing a closed-universe interpretation on the perturbed problem}

\mbox{ }

\noindent
N.B. the above are universes not modes within a particular universe -- 
one has just the one value of $\fE$, $\fE_{\su\sn\si\sv}$.  
Sometimes \cite{DeWitt}, this corresponds to no allowed J, j, sometimes to one and at least sometimes 
to more than one.  
For example, $E_{0, 0} = E_{1, 0}$ for $B = \sqrt{15/2}\hbar^2$, which is perturbatively acceptable 
provided that $A >> B$.  
From this, one can extract partition functions using \cite{LLSM} to take into account that the energies are only known 
perturbatively. 
Hence one can extract a classical notion of entropy and hence of information. 
Also, from knowing the wavefunctions, one can construct mixed states and then the quantum-mechanical 
von Neumann information corresponding to these.

But for records-theoretic purposes, it is information content of {\sl subsystems} and relative 
information between subsystems that look to be more significant quantities.  
These are obtained from solving the subsystem quantum problems (subject to global restrictions such that the 
subsystem energies add up to a fixed energy of the universe \cite{06I}) and then proceeding to compute 
notions of information similar to the above (but now including {\sl relative} information between 
pairs of subsystems that is a quantity of higher relevance as regards records-theoretic approaches).  
This would require considerable further study.

\mbox{ }

\noindent{\large\bf 2.12 Exploiting further correspondences with the rotor problems} 

\mbox{ }

\noindent
The following are available in the rotor literature (in the spherical presentation) and could thus be 
straightforwardly transcribed to the various presentations for this paper's scalefree triangleland problem.
Higher order corrections in $B$ are in the literature for the rotor \cite{Hughes, RPS} and so can be 
transcribed into the relational context [e.g. I was able to write $O(B^4)$ and not $O(B^3)$ in 
(\ref{Diesel}) due to this].  
Alternative variational methods (using the Hellmann--Feynman and hypervirial theorems) appear in 
\cite{RPS}; 
these cover higher order terms too, and can be used to show generally that only even powers of $B$ occur.  
The calculation for the large $B$ regime has both been done \cite{Proprin} and matched to small $B$ 
regime calculations.

Numerical evaluation of eigenvalues was been done by Lamb's \cite{Lamb} continued fraction 
method \cite{Hughes, KuschHughesShirley, PS57} (and otherwise, e.g. \cite{Meyenn, RPS}).  
Myself, I just simply used Maple's \cite{Maple} rkf45 solver alongside the iteration of the method in 
e.g. \cite{Robinett} to locate the eigenvalues, which gives reasonable agreement with the formula 
at the end of SSec 2.10.

The rotor literature also indicates how the $C$ perturbation can be transformed away with a new rotated 
choice of coordinates in which the maths is again that of a $B$ type perturbation (Sec 3 -- the quantum 
application of the trick in Sec I.5).  
While in the laboratory with a rotor one could choose one's axial `$z$' direction to be in the most 
convenient direction, there are various Problem of Time strategy modelling reasons not just to stay in 
these coordinates in our triangleland problem (see the next Sec).

\section{General triple harmonic oscillator problem for scalefree triangleland}  

In the spherical coordinate representation, this takes the form  
\beq
\frac{1}{\mbox{sin}\Theta}\frac{\pa}{\pa\Theta}
\left\{ 
\mbox{sin}\Theta\frac{\pa\Psi}{\pa\Theta}  
\right\} 
+ \frac{1}{\mbox{sin}^2\Theta}\frac{\pa^2\Psi}{\pa\Phi^2} 
+ \{{\cal E} - {\cal A} - {\cal B}\mbox{cos}\Theta - {\cal C}\mbox{sin}\Theta\mbox{cos}\Phi\}\Psi = 0 \mbox{ } ,
\label{BC}
\eeq
where ${\cal A}$, ${\cal B}$, ${\cal C}$ are given by (\ref{ABE}) and 
\beq
{\cal C} = 2C/\hbar^2 \mbox{ } . 
\eeq
This is harder because the angle-dependence of the potential results in nonseparability in these natural 
coordinates, which makes for a useful model of the semiclassical approach to the Problem of Time 
\cite{SemiclIII}.

\subsection{Solution via use rotation/normal modes/adapted basis}

As in \cite{08I}, start again using rotated/normal coordinates at the classical level, or switch to such 
coordinates at the differential equation solving stage, amounting to a choice of basis in which the 
perturbation is in the (new) axial `$z_{\sN}$' direction.
While it can be viewed as before in the new rotated/normal coordinates, nevertheless studying the 
original coordinates' $\Phi$-dependent $\fV$ term remains of interest as it may well be appropriate for 
the {\sl original} coordinates to have mechanical attributes or the heavy--light subsystem distinction 
underlying the semiclassical approach.\footnote{In 
the laboratory, one might likewise not pick the normal coordinates of the rotor--electric field  if 
there is e.g. also a magnetic field that picks out a {\sl different} direction.}   
%
Moreover in that case, being able to proceed further in the rotated/normal coordinates can serve 
as a check on ``standard" procedures in the original coordinates (along the lines suggested in   
\cite{SemiclI}), e.g. as a test of the validity and accuracy of assumptions and approximations made 
in the semiclassical approach.  
This section serves to begin to set up such a check-point.  
Another reason why one might not adopt the normal coordinates as the physically significant ones 
is in order to use J as a trackable non-conserved quantity as regards investigating the semblance 
of dynamics in a fundamentally timeless records theory approach.

There is now an issue in the projection that there is an additional factor due to the change in area in 
moving between each patch of sphere and each corresponding patch of plane.  
Namely, the probability density function on the sphere is $\mbox{sin}\Theta|\Psi(\Theta, \Phi)|^2$ of which the 
$\mbox{sin}\Theta$ pertains to the sphere itself, while the probability density function on the 
stereographic plane is  ${\cal R}/\{1 + {\cal R}^2\}|\Psi({\cal R}, \Phi)|^2$ of which the 
${\cal R}/\{1 + {\cal R}^2\}$ pertains to the stereographic plane itself.  
The very special solution now suffices as an illustration.
The analytic form of its probability density function is
\beq
\mbox{PDF}(\Theta_{\sN}, \Phi_{\sN}) \propto 
\mbox{sin}\Theta_{\sN}\{P_{\sJ}^{\sj}(\mbox{cos}\Theta_{\sN})\}^2
\eeq
and so 
\beq
\mbox{PDF}(\Theta, \Phi) \propto \mbox{sin}\Theta
\left\{
P_{\sJ}^{\sj}
\left(
\frac{    B\mbox{cos}\Theta + C\mbox{sin}\Theta\mbox{cos}\Phi    }{    B_{\sN}    }
\right)
\right\}^2
\eeq
and so 
\beq
\mbox{PDF}({\cal R}, \Phi) \propto \frac{{\cal R}}{1 + {\cal R}^2}
\left\{
P_{\sJ}^{\sj}
\left(
\frac{B\{1 - {\cal R}^2\} + 2C{\cal R}\mbox{cos}\Phi}{B_{\sN}\{1 + {\cal R}^2\}}
\right)
\right\}^2 \mbox{ } .  
\eeq  
Then, in terms of the partial barycentric moments of inertia,
\beq
\mbox{probability density function}(\mI_1, \Phi) \propto 
\left\{
P_{\sJ}^{\sj}
\left(
\frac{B\{\mI - 2\mI_1\} + 2C\sqrt{\mI_1\{\mI - \mI_1\}}\mbox{cos}\Phi}{B_{\sN}\mI}
\right)
\right\}^2 
\eeq
and
\beq
\mbox{PDF}(\mI_2, \Phi) \propto 
\left\{
P_{\sJ}^{\sj}
\left(
\frac{B\{2\mI_2 - \mI\} + 2C\sqrt{\mI_2\{\mI - \mI_2\}}\mbox{cos}\Phi}{B_{\sN}\mI}
\right)
\right\}^2 \mbox{ } .   
\eeq
For (J, j)  = (0, 0), these are the same as for the $C = 0$ case, so there is no need to provide a 
new plot.
However for higher values of (J, j), they are distinct (Fig 4).  

\mbox{ }

\noindent{\footnotesize[{\bf Figure 4} For B = 1 with C = 0.1, then 1 and then 10, I give the following 
triples of plots.
PDF(${\cal R}$, $\Phi$) for (J, j) = (1, 0).
PDF(${\cal R}$, $\Phi$) for (J, j) = (2, 0).
PDF($\mI_1$, $\Phi$) for (J, j) = (1, 0).
PDF($\mI_1$, $\Phi$) for (J, j) = (2, 0), the first of which has an shallower ring hidden inside the 
visible outer ring.]}


\section{Conclusion}

\subsection{Results Summary}

Study of relational particle mechanics is motivated by the absolute versus relative motion debate and the analogy 
between relational particle mechanics and the canonical formulation of General Relativity.  
Specific quantum similarity relational particle mechanics of N = n + 1 particles in 2-d can be constructed due to knowledge in this case 
that the classical configuration space is $\mathbb{CP}^{\sn - 1}$ \cite{TriCl, FORD, Kendall}.  
Likewise, specific similarity relational particle mechanics of N particles in 1-d can be constructed due to the knowledge that in this case 
the classical configuration space is $\mathbb{S}^{\sn - 1}$.  
This paper provides a quantum study of similarity relational particle mechanics with harmonic oscillator 
like potentials, for which I use exact, asymptotic, perturbative and numerical methods. 
Throughout, a mathematical analogy with the linear rigid rotor in a background electric field is useful; 
I then transcribe this from spherical/stereographic plane terms to be in terms of (partial barycentric 
moments of inertia, relative angle) variables and the original similarity relational particle mechanics problem's mass-weighted relative 
Jacobi variables.  
In particular, 

\noindent 1) in the spherical representation there is a constant-potential subcase soluble thereupon 
in terms of spherical harmonics.  

\noindent 2) I then consider relative angle $\Phi$ independent potentials, corresponding to the existence of 
a relative angular momentum type conserved quantity whereby the classical and quantum theory is simplified.    
This permits solution of the large and small stereographic coordinate regimes in terms of Bessel functions, 
and, more accurately, in terms of Laguerre polynomials; the large regime is, moreover, universal rather 
than dependent on the choice of harmonic oscillator like potentials.  

\noindent 3) There are various theoretical reasons to wish to remove the $\Phi$-indepenent restriction -- 
dynamical and quantum mechanical nontriviality and genericity, as well as toy-modelling semiclassical 
and records theory approaches to the Problem of Time in Quantum Gravity.  
Thus I also treat $\Phi$-dependent potentials by a coordinate rotation/normal modes/adapted basis 
construction.  

\noindent 4) In each case, I then interpret these solutions in terms of the underlying mechanical 
variables, including investigation of which triangles formed by the particles are more and less 
quantum-mechanically probable in a given state.

\subsection{Further Extensions}

\noindent The present paper is furthermore useful in that many methods used in it can furthermore be 
used toward solving other concrete relational particle mechanics examples. 
Following the success of identifying the scalefree triangleland multiple harmonic oscillator like potential problem with 
the linear rigid rotor Stark effect, surveying the Quantum Chemistry literature for analogues of some 
of the below may be useful. 

\mbox{ } 

\noindent{\bf  Scalefree 4-stop metroland also has the configuration space $\mathbb{S}^2$.} 

\mbox{ }

\noindent Then the conformal-ordered time-independent Schr\"{o}dinger equation for multiple harmonic 
oscillator potentials is  
\beq
\frac{1}{\mbox{sin}\Theta}\frac{\pa}{\pa\Theta}
\left\{ 
\mbox{sin}\Theta\frac{\pa\Psi}{\pa\Theta} 
\right\} 
+  \frac{1}{\mbox{sin}^2\Theta}\frac{\pa^2\Psi}{\pa\Phi^2} = 
{\cal A} + {\cal B}\mbox{cos}(2\Theta) + {\cal C}\mbox{sin}^2\Theta\mbox{cos}(2\Phi) \mbox{ } , 
\eeq
or, in Legendre variables, 
\beq
\frac{\pa}{\pa X}
\left\{
\{1 - X^2\}\frac{\pa \Psi}{\pa X} 
\right\} 
+ \frac{1}{1 - X^2}\frac{\pa^2\Psi}{\pa\Phi^2} = 
{\cal D} + X^2\{2{\cal B} - {\cal C}\mbox{cos}(2\Phi)\} + {\cal C}\mbox{cos}(2\Phi) \mbox{ } 
\eeq
where 
${\cal A} = 2\{{\fA} - {\fE}\}$, ${\cal B} = 2{\ttB}/\hbar^2$, ${\cal C} = 2{\ttC}$, and     
${\cal D} = {\cal A} - {\cal B}$.

Then for ${\cal B} = {\cal C} = 0$, one obtains the Legendre equation. 
Then for the ${\cal B}$ perturbation, double use of the recurrence relation (\ref{*})\foo{An 
alternative computational scheme is to proceed by the formulae for integrals of products of three 
spherical harmonics in e.g. (\cite{Blum}).}  
gives rise to a $\Delta \mj = 0$, $\Delta \mJ = 0, \pm 2$ `selection rule'.
While, the ${\cal C}$ perturbation has

\noindent
$
\int_{-1}^{1}P_{\sJ}^{\sj \pm 2}(X)\{1 - X^2\}P_{\sJ}^{\sj}(X)\d X
$
contributions, for which double use of the recurrence relation (\ref{**}) gives in this case the 
selection rule $\Delta\mj = \pm 2$, $\Delta \mJ = 0, \pm 2$.   
Note that normal mode trick doesn't work for this problem, so here one has to study the ${\cal C}$ 
perturbation as well as the ${\cal B}$ perturbation.

\mbox{ }

\noindent{\bf Higher N-stop metrolands}

\mbox{ }  

\noindent Here the conformal-ordered time-independent Schr\"{o}dinger equation is, in terms of 
ultraspherical angles,  
\beq
\frac{1}{\prod_{j = 1}^{A - 1}\mbox{sin}^2\Theta_j \mbox{sin}^{\sn - 1 - A}\Theta_{A}}
\frac{\pa}{\pa\Theta_{A}}
\left\{
\mbox{sin}^{\sn _ 1 - A}\Theta_A \frac{\pa\Psi}{\pa\Theta_A}
\right\}
= - {\cal E}\Psi + \sum_{p = 1}^{\sn}{\cal K}_{p}n_{p}^2\Psi  
\eeq
where ${\cal E} = 2\fE/\hbar^2$, ${\cal K}_{\barp} = K_{\barp}/\hbar^2$, $n_{p}$ is the unit vector 
of the embedding Euclidean configuration space ${\cal R}(\mN, 1) = \mathbb{R}^{\sn}$ and 
$\fE$ has been redefined to incorporate the conformal term since (hyper)spheres are of constant Ricci 
scalar.

There is also a highly special constant potential case within the multiple harmonic oscillator like potentials.  
This is now a more complicated sequence of associated Gegenbauer equations as explained in Appendices 
A and B; these are nevertheless also fairly standard and well-documented \cite{AS, GrRy}).  
One can then study perturbations about this.  
Then, if the associated quantum number is not zero, one does not get the Gegenbauer pairings or the 
right weights straight away (see Appendix B).  
Computation of first order perturbations in this case requires the Gegenbauer parameter converting 
recurrence relation $(\ref{GegRec2})$ as well as the polynomial order reducing recurrence relation 
(\ref{GegRec1}), making the calculation somewhat more complicated in this case.  

\mbox{ }

\noindent{\bf Yet further extensions}

\mbox{ }

\noindent As regards (N $>$ 3)-a-gonland's `genuine' $\mathbb{CP}^k$ mathematics (rather than 
$\mathbb{CP}^1$ mathematics that is re-expressible as $\mathbb{S}^2$ mathematics) is required, placing 
this beyond the scope of the present paper.

An alternative type of extension is to keep a given geometry but consider a variety of other potentials, 
e.g. Coulomb-like potentials or a mixture of Coulomb-like and harmonic oscillator like potentials.

\subsection{\bf Applications to classical and quantum geometrodynamics}

The semiclassical approach to the Problem of Time in Quantum Gravity requires nonseparability so that 
the crucial cross-term in (\ref{CruCro}) exists.    
To set up such a model, one could have e.g. two heavy (H) particles and one light (L) one, leading to 
one H relative Jacobi separation and one L one.  
The nonseparability requirement would mean requiring the interpretation that the physics picks out 
$\Theta$ and $\Phi$ variables unaligned with the simplifying normal modes ones (in terms of which there 
{\sl is} separability), and that the potential be $\Phi$-dependent.     
Another way of setting up a semiclassical approach model using the material in this paper is if 
$\Theta$ and $\Phi$ are considered to be H and L respectively (the opposite assignment is impossible 
as sin$^2\Theta$ cannot be $>> 1$).  
The above two situations would ba a significant improvement on the example in \cite{SemiclI} because 
that has no nontrivial linear constraints; it remains to be seen how far such a calculation could be 
taken analytically (and, if needs be, numerically and/or subject to further approximations).   
While, the \cite{SemiclI} example's ability to be solved by usually-unavailable means external to 
the semiclassical approach is retained by the above examples -- it is the work in the present paper. 
Thereby one can assess whether the semiclassical ans\"{a}tze and associated (and any other non-associated 
but unavoidable) approximations are sensible for these new semiclassical approach models.  
To have a {\sl shape--scale aligned} H--L split, which in some ways more closely parallels 
GR Cosmology (H scalefactor versus L inhomogeneities and/or anisotropies), one needs to study 
the Euclidean relational particle mechanics counterpart of this paper \cite{08III}.    
This is also the case if one wishes to obtain a dilational (York-like) internal time model 
\cite{06II, SemiclI}.  

\mbox{ }

As regards building concrete relational particle mechanics examples of the records theory approach to 
Quantum Gravity, a notion of distance on configuration space readily follows \cite{Records} from the 
metrics in \cite{FORD}.
As regards computing a notion of information/negentropy, given an explicitly solved QM, one can (c.f. 
Sec 2.11) build a statistical mechanics from that and extract the negentropy/information.
Moreover, this continues to be the case when the QM is only known perturbatively \cite{LLSM} (with 
correction terms up to the corresponding perturbative order \cite{LLSM}), which is one reason for the 
relevance of the perturbative calculations in this paper.   
Further questions then are: does this paper's model have a notion of information storage? 
(In parallel with \cite{H99}, might a heavy particle passing by -- the signal -- imprint 
separation/motion on the other two particles -- the record? 
\cite{H99} proceeds to study this via path integrals and associated objects such as the influence 
functional; the extent to which these more specialized computations have been carried out for the 
analogous rotor is an interesting question, whose investigation might substantially cut down on how 
much work is needed to complete the parallel calculation to \cite{H99} for the present paper's 
models.)

As regards semblance of dynamics emerging from timeless records theory, the relative angular momentum 
${\cal J}$ that was a conserved quantity for $B = 0$ becomes a changing quantity for $C \neq 0$, so 
tracking and explaining that may be of significance.      
I can track this classically by e.g. computing it at each stage in the rkf45 routine.    
An interesting question then is whether and how this could be tracked quantum-mechanically?  
As regards whether evidence can be found for/against Barbour's conjecture of time capsules \cite{B94II, 
EOT}, the probability density functions plotted in this paper are the right output for addressing that. 
The fairly standard maths I obtain (at least in my simple specific example of harmonic oscillator like 
potentials) suggests that, if time capsules do occur, they ought to be findable also within standard QM 
(for perturbed linear rigid rotors).
However, further scaled triangleland or trihaedronland (= 3-haedronland) work would be necessary as 
regards more specific conjectures Barbour makes about time capsules that were specifically made about 
3-particle Euclidean relational particle mechanics and whether its triple collision or uniform (i.e. 
equilateral) configurations play highly dominant roles (where the probability density function `mist' 
might be highly concentrated).  
For the moment, my simple similarity relational particle mechanics model would not seem to exhibit 
very heavy peaking around its equilateral configuration.

This paper's model is also conceivably an interesting one from the perspective of histories theory and 
as regards the problem of finding (partial) observables for Quantum Gravity.

\mbox{ } 

On the whole, this paper downplays the suggestion \cite{BS89, EOT} that (Barbour's) relationalism 
requires radically different QM theory, though one would need to check further examples (including 
more complicated ones) to be more sure of this conclusion (some of the further examples above can 
be motivated on such grounds).  
Some closed-universe and finite-universe effects are, however, manifest.
As regards the issue \cite{DeWitt, EOT} of whether stationary quantum universes have single or multiple 
states, I comment that this paper's models do exhibit some degenerate states (both among simple exact 
solutions and perturbatively to second order); however one needs a much more extensive study of 
relational particle mechanics model universes before one can begin to say whether these are, however, 
non-generic.
There is also a certain amount of tension between results obtained by reduced quantization as in this 
paper and by Dirac-like quantization as in \cite{06I}, though I postpone discussion of this to 
\cite{08III} (one of the problems being that my method of Dirac-like quantization in \cite{06I} for 
Euclidean relational particle mechanics does not directly extend to similarity relational particle 
mechanics, so that checks between the two methods and lessons drawn from them are best left to the 
Euclidean relational particle mechanics arena).

Treating cases with more than three degrees of freedom will be necessary to get a grip of some aspects.
This is the case firstly for investigation whether there are kinetic effects of the kind that Barbour 
conjectures \cite{B94II, EOT} as regards a semblance of time arising from timeless configurations.
All of the $\mathbb{S}^{\sn - 1}$ being conformally flat, one would need to extend to such as the 
non-conformally flat $\mathbb{CP}^2$ so that effects that are {\sl irreducibly} kinetic occur 
(rather than cases for which a kinetic and potential re-definition leaves one with a flat kinetic 
term in the end by passing the conformal factor into a re-defined potential).  
Secondly, for (N $\geq 4$)-stop metroland, the relative angle dependence comes from writing the 
potential in coordinates dictated by the kinetic term, and no longer in such a way that the standard 
rotation to normal coordinates removes this complication.  
A third such point, which constitutes an interesting further investigation in its own right, 
is that N-particle models also have a robustness application: is the QM of N -- 1 
particles stable to the inclusion of a further particle?  
This parallels the situation of whether Taub space is stable within the mixmaster solution in 
minisuperspace Quantum Cosmology (found to be unstable in \cite{KR89}).
Fourthly, upgrading to models with $>$ 3 particles is likely to improve one's capacity to 
use particle clumps to model inhomogeneities.  
A final question that such models can be used to investigate is whether the conformal ordering 
succeeds in avoiding conflict with other necessary technical conditions.  

\mbox{ }  

\noindent{\bf Acknowledgments}

\mbox{ } 

\noindent I thank Dr Julian Barbour, Professor Chris Isham and Miss Anne Franzen for discussions, 
the anonymous Referees for comments, the organizers of ``Space and Time 100 Years after Minkowski" 
Conference at Bad Honnef, Germany, the Perimeter Institute, the Spinoza Institute and Queen Mary's 
Relativity Group for invitations to speak and for hospitality and Dr Julian Barbour also for hospitality. 
Peterhouse for funding in 2006-08. 
Professors Malcolm MacCallum, Gary Gibbons, Don Page, Reza Tavakol and Jonathan Halliwell,
and Dr's Julian Barbour and Fay Dowker, for support in the furthering of my career.   
My Wife for help with assembling the figures, and my Wife, Alicia, Amelia, Beth, Emma, Emilie, Emily, 
Joshua, Luke, Simeon and Will for keeping my spirits up.    
 
\mbox{ }

\noindent{\bf \large Appendix A}

\mbox{ }

\noindent  
Following from App I.A.2, the energy constraint gives, in conformal ordering, the time-independent 
Schr\"{o}dinger equation
\beq
\hat{H}\Psi \equiv
- \frac{\hbar^2}{2}
\left\{
\frac{1}{\sqrt{{\cal M}}}
\frac{\pa}{\pa{\cal Q}^{\sfA}}
\left\{
\sqrt{{\cal M}} {\cal N}^{\sfA\sfB}\frac{\pa \Psi}{\pa{\cal Q}^{\sfB}} 
\right\} 
- \frac{k - 2}{4\{k - 1\}}\mbox{Ric}({\cal M})\Psi
\right\}
+ \fV\Psi = \fE\Psi \mbox{ } .  
\eeq

\mbox{ }

\noindent{\bf A.1 Scalefree N-stop metroland} 

\mbox{ }

\noindent
Here, by p 1269-70 of \cite{Isham84}, an appropriate finite subalgebra acting on the corresponding 
cotangent space is then SO$(\mn\d) \mbox{\textcircled{S}} \mathbb{R}^{\sn\sd}$, where \textcircled{S} 
stands for semidirect product.      
This can be considered to be generated by angular momenta $J_i$ and coordinates $u_i$ such that 
$\sum_iu_i^2 = 1$.

Now, the Laplacian corresponding to line element (I.20) is 
\beq
D^2 = \prod_{j = 1}^{\sn\sd - 2}\mbox{sin}^{j - \sn\sd + 1}\Theta_j \frac{\pa}{\pa\Theta_a}
\left\{
\frac{\prod_{j = 1}^{\sn\sd - 2}\mbox{sin}^{\sn\sd - 1 - j}\Theta_j}
{\prod_{i = 1}^{a - 1}\mbox{sin}^{2}\Theta_i}\frac{\pa}{\pa\Theta_a}  
\right\} 
= \frac{1}{\mbox{sin}^{\sn\sd - 1 - A}\Theta_A\prod_{i = 1}^{A - 1}\mbox{sin}^2\Theta_i}
\frac{\pa}{\pa\Theta_A}\left\{\mbox{sin}^{\sn\sd - 1 - A}\Theta_A\frac{\pa}{\pa\Theta_A}\right\} \mbox{ } .
\eeq
This situation applies to preshape space and for shape space \cite{Kendall, FORD}.  
Hence QM on preshape space for N particles in dimension d has the time-independent Schr\"{o}dinger 
equation
\beq
-\frac{\hbar^2}{2}\frac{1}{\mbox{sin}^{\sn\sd - 1 - A}\Theta_A\prod_{i = 1}^{A - 1}\mbox{sin}^2\Theta_i}
\frac{\pa}{\pa\Theta_A}\left\{\mbox{sin}^{\sn\sd - 1 - A}\Theta_A\frac{\pa\Psi}{\pa\Theta_A}\right\}
+ \fV\Psi = \fE\Psi \mbox{ } 
\label{TISE1}
\eeq
(where the energy has been displaced by a constant term from the constant-curvature contribution to 
the conformal ordering).
The d = 1 case of this is, additionally, the time-independent Schr\"{o}dinger equation for scalefree 
N-stop metroland, i.e. eq. (\ref{TISE2}).
These Hamiltonians involve suitable quantum operators (see e.g. p 160 of \cite{RS} for an account of the 
properties of the constituent Laplacian operator on $\mathbb{S}^{\sn - 1}$, see also \cite{Norway, 
ThisS2}).

For constant potential, this time-independent Schr\"{o}dinger equation is an equation of form 
$D^2\Psi = \Lambda\Psi$.  
Then the separation ansatz $\Psi = \prod_{\barp = 1}^{\sn - 1}\psi_{\barp}(\Theta_{\barp})$ yields the 
simple harmonic motion equation for $\Theta_{\sn - 1}$ and n -- 2 equations of form 
\beq
\{1 - X^2_{\sn - \sp}\} \frac{\d^2\psi_{\sn - \sp}}{\d X_{\sn - \sp}^2} - 
\{\mp - 1\}  X_{\sn - \sp}\frac{\d\psi_{\sn - \sp}}{\d X_{\sn - \sp}} + 
\mj_{\sp - 1}\{\mj_{\sp - 1} + \mp - 2\}\psi_{\sn - \sp} - 
\frac{\mj_{\sp - 2}\{\mj_{\sp - 2} + \mp - 3\}}{1 - X^2_{\sn - \sp}}\psi_{\sn - \sp} = 0
\eeq
under the transformations $X_{\hat{\sp}} = \mbox{cos}\Theta_{\hat{\sp}}$, $\hat{\mp} =  1$ to n -- 2.  
These are associated Gegenbauer equations (\ref{AssocGeg}) with parameter $\lambda_{\sp} = \{p - 2\}/2$, 
where the integers $j_{\sp - 1}, j_{\sp - 2} \in \mathbb{N}_0$ are picked out as eigenvalues, so that 
the \{n -- p\}th equation is solved by $C^{\sj_{\tp - 2}}_{\sj_{\tp - 1}}(\mbox{cos}\Theta_{\sn - \sp}; 
\{\mp - 2\}/2)$.  
Then one gets a sequence of integer quantum numbers beginning with the familiar $|\mj_1| \leq \mj_2$.
%

\vspace{2in}

\noindent{\bf A.2 Scalefree N-a-gonland}

\mbox{ } 

\noindent
In this case, as regards kinematic quantization for scalefree N-a-gonland, use that $\fS(\mbox{N}, 2) = 
\mathbb{CP}^{\sn - 1} = $SU$(\mn)/$U$(\mn - 1)$, which is of the general form $\fQ = G/H$ considered in 
\cite{Isham84}.  
Thus a suitable finite algebra acting on the corresponding cotangent space is 
SU$(\mn) \mbox{\textcircled{S}} \mathbb{R}^{2\sn}$.

The Fubini--Study Laplacian for $\fS$(N, 2) is then
\beq
D^2 = \frac{\{1 + ||{\cal R}||^2\}^{2n - 2}}{\prod_{\barp = 1}^{\sn - 1}{\cal R}_{\barp}}
\left\{
\frac{\pa}{\pa {\cal R}_{\barp}}
\left\{
\frac{\prod_{\barp = 1}^{\sn - 1}{\cal R}_{\barp}}{\{1 + ||{\cal R}||^2\}^{2\sn - 3}}
\{\delta^{\barp\barq} + {\cal R}^{\barp}{\cal R}^{\barq}\}
\frac{\pa}{\pa{\cal R}_{\barq}}
\right\} + 
\frac{\pa}{\pa {\Theta}_{\tip}}
\left\{
\frac{\prod_{\barp = 1}^{n - 1}{\cal R}_{\barp}}{\{1 + ||{\cal R}||^2\}^{2\sn - 3}}
\left\{
\frac{\delta^{\tip\tiq}}{{\cal R}_{\barp}^2} + {1|}^{\tip\tiq}
\right\}
\frac{\pa}{\pa{\Theta}_{\tiq}}
\right\}  
\right\} \mbox{ } , 
\label{Lapla} 
\eeq
so that the conformal-ordered time-independent Schr\"{o}dinger equation for the 2-d shape space of N 
particles is (\ref{TISE3}).

\mbox{ }

\noindent{\bf A.3  Scalefree triangleland case}

\mbox{ }

\noindent In the tilded banal conformal representation, the (3, 2) case's Laplacian in the flat 
coordinates (${\cal R}$, $\Phi$) takes the familiar form  
\beq
D^2 = \frac{1}{{\cal R}}\frac{\pa}{\pa{\cal R}}
\left\{
{\cal R}\frac{\pa }{\pa{\cal R}}
\right\} 
+ \frac{1}{{\cal R}^2}\frac{\pa^2}{\pa\Phi^2} \mbox{ } .  
\eeq
While, in the barred banal conformal representation in spherical coordinates ($\Theta$, $\Phi$), it 
takes the also-familiar form 
\beq
D^2 = \frac{1}{\mbox{sin}\Theta}\frac{\pa}{\pa\Theta}
\left\{
\mbox{sin}\Theta\frac{\pa\Psi}{\pa\Theta}
\right\} 
+ \frac{1}{\mbox{sin}^2\Theta}\frac{\pa^2  }{\pa\Phi^2} \mbox{ } .  
\eeq
While I provide a parallel scheme for general $\mathbb{CP}^{k}$ above, it is the $\mathbb{CP}^{1} = 
\mathbb{S}^{2}$ version actually used in this paper's calculations for which I currently have 
guarantees of good behaviour as an operator.   

\mbox{ }  

\noindent{\bf\large Appendix B: Special functions and mappings of ordinary differential equations}

\mbox{ }

\noindent The Bessel equation of order p,
\beq
v^2w_{vv} + vw_{v} +\{v^2 - \mp^2\}w = 0 \mbox{ } ,  
\label{Bess}
\eeq
is solved by the Bessel functions.
I denote Bessel functions of the first kind by $J_{\sp}(v)$.

The associated Laguerre equation, 
\beq
xy_{xx} + \{\alpha + 1 - x\}y_x + \mn y = 0 \mbox{ } , 
\label{Laguerre}
\eeq  
is solved by the associated Laguerre polynomials $L_{\sn}^{\alpha}(x)$ (and unbounded second solutions).  
The 2-d quantum isotropic harmonic oscillator's radial equation for a particle of mass $\mu$ and 
classical oscillator frequency $\omega$, 
\beq
-\frac{\hbar^2}{2\mu}
\left\{
R_{rr} + \frac{R_r}{r} + \frac{\nm^2R}{r^2}
\right\} 
+ \frac{\mu\omega^2r^2R}{2} = E R \mbox{ } , 
\label{isoho}
\eeq
maps to the associated Laguerre equation under the asymptotically-motivated transformations 
\beq
R = 
\left\{
{\hbar x}/{\mu\omega}
\right\}^{\frac{|\tm|}{2}}
\mbox{e}^{- x/2}y(x) \mbox{ } , \mbox{ }  
x = {\mu\omega r^2}/{\hbar}
\label{isohotrans}
\eeq
and so is solved by 
\beq
R \propto r^{|\sm|}\mbox{e}^{\mu\omega r^2/2\hbar}L_{\sr}^{|\sm|}
\left(
\mu\omega r^2/{\hbar}
\right)
\label{isohosoln}
\eeq
corresponding to the discrete energies $E = \{|\nm| + 2\nr + 1\}\hbar\omega$ for 
radial quantum number $\nr \in \mathbb{N}_0$ \cite{Schwinger, Robinett}.

The associated Legendre equation
\beq
\{1 - X^2\}Y_{XX} - 2XY_X + \{\mJ\{\mJ + 1\} - \mj^2\{1 - X^2\}^{-1}\}Y = 0 \mbox{ } 
\label{Leg}
\eeq
(which is equivalent to the equation
\beq
\mbox{sin}^{-1}\Theta\{\mbox{sin}\Theta Y_{\Theta}\}_{\Theta} + 
\{\nJ\{\nJ + 1\} - \nj^2\mbox{sin}^{-2}\Theta\}Y = 0 \mbox{ } 
\eeq
under the transformation $X = \mbox{cos}\Theta$), is solved by the associated Legendre functions 
$P_{\sJ}^{|\sj|}(X)$ (and unbounded second solutions), for J $\in \mathbb{N}_0$, j $\in \mathbb{Z}$, 
$|\mj| \leq$ J.
We use the standard convention that
\beq 
P^{\sj}_{\sJ}(X) = \{-1\}^{\sj}\{1 - X^2\}^{\frac{\sj}{2}}\frac{\d^{\sj}}{\d X^{\sj}}
\left\{
\frac{1}{2^{\sJ}\mJ !}\frac{\d^{\sJ}}{\d X^{\sJ}}\{X^2 - 1\}^{\sJ}
\right\} \mbox{ } , 
\eeq
whereupon 
\beq
\left\{ 
\sqrt{    \frac{2\mJ + 1}{2}  \frac{    \{\mJ - |\mj|\}!    }{    \{\mJ + |\mj|\}!    }    }
P^{|\sj|}_{\sJ}(X)
\right\}
\label{orthog}
\eeq
is a complete set of orthonormal functions for $X \in$ [--1, 1].  
We also require the recurrence relations \cite{GrRy, AS} 
\beq
XP^{|\sj|}_{\sJ}(X) = 
\frac{\{\mJ - |\mj| + 1\}P^{|\sj|}_{\sJ + 1}(X) + \{\mJ + |\mj|\}P^{|\sj|}_{\sJ - 1}(X)}{2\mJ + 1} 
\mbox{ } , 
\label{*}
\eeq
\beq
\sqrt{1 - X^2}P_{\sJ}^{\sj - 1} = \frac{P^{\sj}_{\sJ - 1} - P^{\sj}_{\sJ + 1}}{2\mJ + 1} \mbox{ } .  
\label{**}
\eeq

\mbox{ }

The Gegenbauer equation 
\beq
\{1 - X^2\} Y_{XX} - \{2\lambda + 1\}XY_X + \mJ\{\mJ + 2\lambda\}Y = 0 
\label{Geg}
\eeq
is solved boundedly by the Gegenbauer Polynomials $C_{\sJ}(X;\lambda)$.    
Normalization for these is provided in e.g. \cite{AS, GrRy}; the weight function is 
$\{1 - X^2\}^{\lambda - \frac{1}{2}}$ between equal-$\lambda$ Gegenbauer polynomials.  
These furthermore obey the recurrence relations \cite{AS, GrRy}
\beq
XC_{\sJ}(X;\lambda) = \frac{  \{\mJ + 1\}C_{\sJ + 1}(X; \lambda) + 
\{2\lambda + \mJ - 1\}C_{\sJ - 1}(X; \lambda)}{2\{\mJ + \lambda\}} \mbox{ } , 
\label{GegRec1}
\eeq
\beq
C_{\sJ + 1}(X; \lambda) = \frac{\lambda\{C_{\sJ + 1}(X; \lambda + 1) - C_{\sJ - 1}(X; \lambda + 1)\}}
                            {\mJ + \lambda + 1}
\label{GegRec2} \mbox{ } .  
\eeq
The associated Gegenbauer equation 
\beq
\{1 - X^2\}Y_{XX} - \{2\lambda + 1\}XY_X + \mJ\{\mJ + 2\lambda\}Y - 
\mj\{\mj + 2\lambda - 1\}\{1 - X^2\}^{-2}Y = 0 
\label{AssocGeg}
\eeq
is solved boundedly by the associated Gegenbauer functions $C_{\sJ}^{\sj}(X; \lambda)$.  
These are re-expressible in terms of Gegenbauer polynomials via \cite{Norway} 
\beq
C_{\sJ}^{\sj}(X; \lambda) \propto \{1 - X^2\}^{\frac{\sj}{2}}C_{\sJ - \sj}(X; \lambda/2 - 1 + \mj) \mbox{ } . 
\label{conv}
\eeq
With this conversion, the recurrence relations between Gegenbauer polynomials (\ref{GegRec1}, 
\ref{GegRec2}) turn out to suffice for this paper.  
For $\lambda = 1/2$, (\ref{Geg}) is the Legendre equation solved by $P_{\sJ}(X) = C_{\sJ}(X;1/2)$, 
(\ref{GegRec1}) becomes (\ref{*}), and (\ref{AssocGeg}) becomes the associated Legendre equation 
(\ref{Leg}) solved by $P^{-\sj}_{\sJ} \propto C^{\sj}_{\sJ}(X; 1/2)$, by (\ref{conv}) and \cite{AS} 
\beq
C_{\sJ}(X;\lambda) \propto \{X^2 - 1\}^{\frac{1}{4} - 
\frac{\lambda}{2}}P^{\frac{1}{2} - \lambda}_{\sJ + \lambda - \frac{1}{2}}(X) \mbox{ } .  
\eeq



\begin{thebibliography}{99}
\bibitem{Maple}               This was done using Maple 10 to 12.
\end{thebibliography}
\end{document}